\documentclass[twocolumn,amsmath,trackchanges]{aastex702}
\usepackage[flushleft]{threeparttable}

\begin{document}
\defcitealias{2024MNRAS.530.1866S}{S24}
\title{Particle Acceleration in Stratified 2D MRI Turbulence: From Reconnection to First- and Second-Order Fermi and Shear Acceleration}

\author[orcid=0009-0009-7362-1311,gname=Astor, sname='Sandoval Parra']{Astor Sandoval}
\affiliation{Millennium Nucleus on Transversal Research and Technology to Explore Supermassive Black Holes (TITANS), Chile}
\affiliation{Departamento de Ciencias, Facultad de Artes Liberales, Universidad Adolfo Ibáñez, Av. Padre Hurtado 750, Viña del Mar, Chile}
\email[show]{astor.sandoval@edu.uai.cl}  

\author[orcid=0000-0003-2928-6412]{Mario Riquelme} 
%\altaffiliation{Las Campanas Observatory}
\affiliation{Departamento de F\'isica, Facultad de Ciencias F\'isicas y Matem\'aticas, Universidad de Chile, Santiago, RM 8370449, Chile}
\email{marioriquelme@uchile.cl}

\author[orcid=0000-0003-1965-3346]{Jorge Cuadra}
\affiliation{Millennium Nucleus on Transversal Research and Technology to Explore Supermassive Black Holes (TITANS), Chile}
\affiliation{Departamento de Ciencias, Facultad de Artes Liberales, Universidad Adolfo Ibáñez, Av. Padre Hurtado 750, Viña del Mar, Chile}
\email{jorge.cuadra@uai.cl}

%\collaboration{all}{The Terra Mater collaboration}

\begin{abstract}

Low-luminosity accretion disks around black holes are weakly collisional, potentially enabling non-thermal particle acceleration. Local two-dimensional particle-in-cell simulations of the stratified magnetorotational instability (MRI) in pair plasmas by \citet{2024MNRAS.530.1866S} (hereafter S24) revealed non-thermal populations with maximum energies proportional to the scale-separation ratio $\omega_{c,0}/\Omega_0$, where $\omega_{c,0}$ and $\Omega_0$ are the initial particle cyclotron and Keplerian frequencies, respectively. We identify the acceleration mechanisms involved and quantify their scale-separation dependence. Acceleration occurs primarily within large-scale current sheets associated with an MRI dynamo. Particles are initially injected by non-ideal electric fields during reconnection driven by plasmoid splitting, attaining energies independently of scale separation. They subsequently undergo rapid first-order Fermi-like acceleration occurring within converging flows between newly formed, separating plasmoids. During this stage, energies grow at a rate proportional to the instantaneous non-relativistic cyclotron frequency, independently of $\omega_{c,0}/\Omega_0$. This fast acceleration rate implies a maximum energy proportional to $\omega_{c,0}/\Omega_0$. Particles then continue interacting repeatedly with plasmoids, undergoing slower, predominantly second-order Fermi acceleration, with a subdominant contribution from shear acceleration. This stage increases particle energies by only a factor of {\it a few} over several orbits, regardless of $\omega_{c,0}/\Omega_0$. The combination of these three stages implies a maximum energy proportional to $\omega_{c,0}/\Omega_0$, explaining the scaling found by S24 and suggesting that the MRI may accelerate particles to ultrarelativistic energies in low-luminosity disks. Testing these results in three dimensions remains an important next step.

\end{abstract}



\section{Introduction}

Accretion disks are ubiquitous in astrophysics, spanning a wide range of mass scales \citep{1981ARA&A..19..137P,2012arXiv1203.6851M,2013LRR....16....1A}. In these systems, the magnetorotational instability \citep[MRI;][]{1991ApJ...376..214B,1998RvMP...70....1B} plays a central role in driving the outward angular momentum transport. Most of our understanding of the MRI-driven turbulence comes from magnetohydrodynamic (MHD) simulations, which treat the plasma as a fully collisional fluid. However, in low-luminosity accretion disks particle collisions are rare, breaking local thermodynamic equilibrium and rendering the plasma effectively collisionless. In this regime, non-thermal processes can take place, such as electron and ion temperature decoupling, the appearance of pressure anisotropies, and the acceleration of particles to relativistic energies. 

Collisionless disks are believed to occur in the low/hard state of X-ray binaries  \citep{1997ApJ...489..865E}, as well as around low-luminosity black holes such as Sagittarius A$^*$ (Sgr $A^*$) and M87 \citep{1997ApJ...490..605M,2003ApJ...598..301Y,2018ApJ...864..126R}. High-energy $\gamma$-ray observations suggest that Sgr A$^*$ can accelerate protons to energies up to the PeV \citep{2006ApJ...636..777A,2016Natur.531..476H,2018A&A...612A...9H} and M87 \citep{2006Sci...314.1424A,2012A&A...544A..96A,2012ApJ...746..141A} could accelerate protons up to hundreds of TeV \citep{2025ApJ...990..170X} or electrons up to hundreds of GeV \citep{2026ApJ...997..151M} during the quiescent state. Similarly, the detection of high-energy neutrinos
suggests that accretion onto extragalactic supermassive black holes, such as NGC 1068 \citep{2022Sci...378..538I} may also accelerate ions to tens or hundreds of TeV \citep{2023ApJ...956....8F,2026arXiv260215644E,2026arXiv260401222D}. 

In addition, episodic flaring activity \citep{2018A&A...618L..10G} may drive the acceleration of electrons up to hundreds of GeV in Sgr A$^*$ \citep{2017MNRAS.468.2447P} and up to TeV in M87 \citep{2026ApJ...997..151M}. However, despite these observational suggestions of efficient particle acceleration in disks, the precise location and mechanisms of particle acceleration remain an open problem. Proposed processes include:
\begin{itemize} 
    \item  Internal shocks \citep{2011ApJ...743...47B,2017MNRAS.465.1409L,2022MNRAS.514.1940D},
    \item Magnetic reconnection \citep{2012ApJ...755...50R,2013ApJ...773..118H,2015PhRvL.114f1101H,2018ApJ...859..149I,2020ApJ...900..100R,2022ApJ...938...86B,2024PhRvL.133d5202B,2023PhRvR...5d3023Z,2024MNRAS.530.1866S,2025ApJ...980..255Y}, and 
    \item Turbulent and shear acceleration \citep{2016ApJ...822...88K,2019MNRAS.485..163K,2021MNRAS.506.1128S}.
\end{itemize}

In this work, we investigate the potential role of the MRI in collisionless disks in particle acceleration using particle-in-cell (PIC) simulations. Previous PIC studies have shown that the MRI can indeed accelerate particles \citep{2012ApJ...755...50R,2013ApJ...773..118H,2015PhRvL.114f1101H,2016ApJ...822...88K,2018ApJ...859..149I,2022ApJ...938...86B,2024PhRvL.133d5202B,2024MNRAS.530.1866S,2025ApJ...982L..28G}. In particular, \citet{2024PhRvL.133d5202B} demonstrated that this acceleration is a robust process that occurs during the saturated regime of the MRI turbulence, while \citet{2025ApJ...982L..28G} showed that this acceleration can be equally efficient for ions and electrons. Furthermore, \citet[][hereafter \citetalias{2024MNRAS.530.1866S}]{2024MNRAS.530.1866S} employed for first time PIC simulations of the MRI in vertically stratified disk, finding that the stratification plays a critical role in regulating the plasma conditions under which the acceleration takes place. For instance, the presence of disk expansion and plasma outflows in stratified simulations gives rise to a large scale magnetic field produced by a dynamo process, which at the same time increases the plasma magnetization. 
Interestingly, the results of \citetalias{2024MNRAS.530.1866S} show that the maximum energy of the accelerated particles is proportional to the scale separation ratio of the simulation $\omega_{c,0}/\Omega_0$, where $\Omega_0$ and $\omega_{c,0}$ are the Keplerian frequency of the orbiting plasma and the initial cyclotron frequency of the particles, respectively. Although the resulting acceleration parameters—spectral index and maximum energy—are consistent with magnetic reconnection predictions, several key questions remain open:
\begin{enumerate}
\item Is magnetic reconnection the sole driver, or do other mechanisms such as shear acceleration \citep{1981SvAL....7..352B,1988ApJ...331L..91E,2006ApJ...652.1044R,2013ApJ...767L..16O,2019PhRvD..99h3006L,2021MNRAS.506.1128S} and/or second order Fermi \citep{2005A&A...441..845D,2011ApJ...735..102K,2012PhRvL.108x1102K,2012SSRv..173..557L,2014ASPC..488....8D,2016MNRAS.463.4331D,2020ApJ...899..151K,2022JPlPh..88a9014U} also contribute?
\item What is the role played by the dynamo field in the acceleration process?
%\item Where in the disk does the acceleration occur—near the mid-plane or in the corona? %Is it 
\item What limits the maximum attainable particle energy-is it limited by the size of the reconnecting region \citep[see e.g.,][]{2016ApJ...816L...8W} or by the acceleration time \citep[e.g.,][]{2021ApJ...922..261Z}? 
\end{enumerate}

To address these questions, we use the 2D stratified shearing-box particle-in-cell (PIC) simulations presented in \citetalias{2024MNRAS.530.1866S}, in which we focus on the sub-relativistic regime relevant to inner black hole accretion disks. For computational convenience, our simulations employ a pair plasma (equal electron and ion masses). Because the inertia in realistic accretion disks is dominated by ions—and since MRI-driven turbulence decays first at ion kinetic scales—we expect that the acceleration mechanisms identified here are likely mostly relevant for ion dynamics.

This paper is organized as follows. Section \ref{sec:sim_setup} describes the numerical method and simulation setup. Section \ref{sec:turbulence} presents key properties of the stratified MRI turbulence. Section \ref{sec:acc28} analyzes particle injection and acceleration during the MRI evolution for simulation with scale separation $\omega_{c,0}/\Omega_0=28$, and Section \ref{sec:scale_sep} explores how the acceleration mechanism depends on $\omega_{c,0}/\Omega_0$. Finally, our conclusions are presented in Section \ref{sec:conc}.

\section{Simulation Setup} \label{sec:sim_setup}

We perform two-dimensional PIC simulations using the \textsc{TRISTAN-MP} code  \citep{buneman1993computer,2005AIPC..801..345S} in two dimensions to model the stratified kinetic MRI using a pair plasma. 
The simulations employ the local shearing-box approximation \citep{1995ApJ...440..742H}, representing a small, stratified patch of an accretion disk.

The computational domain uses cartesian coordinates: $x$ (radial), $y$ (azimuthal), and $z$ (vertical). The frame rotates at angular frequency $\boldsymbol{\Omega_0} = \Omega_0 \hat{z}$ and includes a shear velocity profile
\begin{equation}
    \mathbf{v}_s=-\frac{3}{2}\Omega_0x\hat{y},
\end{equation}
arising from differential rotation in a Keplerian disk. To preserve vertical stratification, we include the gravitational force from the central black hole:
\begin{equation}
    \mathbf{F}_z=-m\Omega_0^2z\hat{z},
\end{equation}
where $m$ is each particle's mass and $z$ the vertical displacement from the mid-plane.

The initial density profile assumes an isothermal disk in hydrostatic equilibrium:
\begin{equation}
        n(z) = n_0 \exp\!\left(-\frac{z^2}{H_0^2}\right), \label{eq:init_dens}
\end{equation}
where $n_0$ is the mid-plane density and $H_0 = (2k_B T_0/m\Omega_0^2)^{1/2}$ the vertical scale height for the initial temperature $T_0$.

The initial conditions in all our simulations are:
\begin{itemize}
    \item Temperature: $k_BT_0=5\times10^{-3}\,mc^2$, where $k_B$ is the Boltzmann constant and $c$ is the speed of light.
    \item Uniform vertical magnetic field $\mathbf{B}_0=B_0\hat{z}$,
    \item Mid-plane plasma beta: $\beta_0=\frac{n_0k_BT_0}{B_0^2/8\pi}=100$
\end{itemize}

Table \ref{tab:param} summarizes the simulations parameters, including box size, grid resolution, and scale-separation ratio $\omega_{c,0}/\Omega_0$, where $\omega_{c,0}=|q|B_0/mc$, and $|q|$ is the magnitude of each particle's charge. This parameter essentially determines how well separated the microphysical and MHD scales are. In real accretion disk, the scale-separation ratio is many order of magnitude larger than we can currently solve in numerical simulations (e.g., considering a magnetic field of ${\sim 10\, \text{G}}$ \citep{2024ApJ...964L..26E} for Sgr A$^*$ implies that for protons $\omega_{c,0}/\Omega_0\approx 10^{10}$ at tens of gravitational radius). Therefore, we explore its impact on the acceleration process by treating it as an additional physical parameter that can be varied.

\begin{deluxetable*}{ccccc}
%\digitalasset
\tablewidth{0pt}
\tablecaption{Simulation Parameters. \label{tab:param}}
\tablehead{ 
\colhead{Run} & \colhead{ST2D-28} & \colhead{ST2D-14} & \colhead{ST2D-7} & \colhead{ST2D-3.5}
}  
\startdata
$\omega_{c,0}/\Omega_0$ & 28 & 14 & 7 & 3.5 \\ 
$L_x$ [$2\pi v_{A,0}/\Omega_0$]  & 37 & 49 & 46 & 48 \\
$L_z$ [$2\pi v_{A,0}/\Omega_0$]  & 100 & 95 & 93 & 96 \\
$\Delta$ [$c/\omega_{p,0}$]  & 0.35 & 0.35 & 0.35 & 0.35 \\
$N_{\mathrm{ppc}}$  & 400 & 300 & 200 & 200 \\
$\Delta t$ [$\Delta/c$] & 0.45 & 0.45 & 0.45 & 0.225 \\
t$_{\text{end}}$ [$2\pi/\Omega_0$] & 4.91 & 5.01 & 5.8 & 4.03\\
\enddata
\tablecomments{ Summary of the initial simulation parameters. The scale-separation ratio $\omega_{c,0}/\Omega_0$ is defined using the initial particle cyclotron frequency ${\omega_{c,0} = |q|B_0/mc}$. The box sizes along the axes ($L_x$ and $L_y$) are given in units of the fastest-growing MRI wavelength, ${\lambda_{\mathrm{MRI}} = 2\pi v_{A,0}/\Omega_0}$. The grid spacing $\Delta$ (equal in both dimensions) is expressed in terms of the initial plasma skin depth, ${c/\omega_{p,0} = c/(4\pi n_0 q^2/m)^{1/2}}$. $N_{\mathrm{ppc}}$ gives the number of particles per cell in the mid-plane (counting both ions and electrons), the time step $\Delta t$ is given in units of $\Delta/c$, and the simulation duration t$_{\text{end}}$ in orbital periods $(2\pi/\Omega_0)$. As for the names of the runs, we are keeping the same ones used in \citetalias{2024MNRAS.530.1866S} to ease comparison.}
\end{deluxetable*}

\subsection{Solved equations}

In order to implement the shearing-box approximation, we use the shearing coordinates framework described in~\citet{2012ApJ...755...50R}. The use of shearing coordinates modifies the Maxwell's equations, and the evolution of particles momentum and position. The full 3D modified Maxwell's equations are \citepalias[see, e.g.,][]{2024MNRAS.530.1866S}
\begin{align}
    \frac{\partial \mathbf{E}}{\partial t} &= c\nabla \times \mathbf{B} - 4\pi \mathbf{J} %-\frac{3}{2}\Omega_0 E_x\hat{y}
    - \frac{3}{2}\Omega_0 tc \frac{\partial \mathbf{B}}{\partial y} \times \hat{x}, \label{eq:Ampere_mod-full}\\
    \frac{\partial \mathbf{B}}{\partial t} &= -c\nabla \times \mathbf{E}
    - \frac{3}{2}\Omega_0 B_x \hat{y}
    + \frac{3}{2}\Omega_0t c\, \frac{\partial \mathbf{E}}{\partial y} \times \hat{x}, \label{eq:Faraday_mod-full}
\end{align}
where $\mathbf{J}$ is the current density and $\mathbf{E}$ and $\mathbf{B}$ are the electric and magnetic fields, respectively.

Equations \ref{eq:Ampere_mod-full} and \ref{eq:Faraday_mod-full} correctly describe the MHD evolution of the fields as long as $v_{A,0} \ll c$, as explained in \citetalias{2024MNRAS.530.1866S}.

In the rotating shearing frame, particles experience the Lorentz force, vertical gravity, and the Coriolis force. The evolution of the particle momentum $\mathbf{p} = (p_x, p_y, p_z)$ is governed by \citepalias[see][]{2024MNRAS.530.1866S}
\begin{equation}
    \frac{d\mathbf{p}}{dt} = q\!\left(\mathbf{E} + \frac{\mathbf{v}}{c} \times \mathbf{B}\right) + 2\Omega_0 p_y \hat{x} - \frac{1}{2}\Omega_0 p_x \hat{y} - m\Omega_0^2 z \hat{z}, \label{eq:momentum_evol}
\end{equation}
where $\mathbf{v} = \mathbf{p}/(\gamma m) = (v_x, v_y, v_z)$ is the particle velocity, and $\gamma$ is the particle Lorentz factor. 

Finally, the particle position $\mathbf{r} = (x, y, z)$ evolves according to
\begin{equation}
    \frac{d\mathbf{r}}{dt} = \mathbf{v} + \frac{3}{2}\Omega_0 t v_x \hat{y}, \label{eq:position_evol}
\end{equation}
where the last term accounts for the velocity gradient correction in the shearing frame. Since our 2D simulations assume axisymmetry ($y$-independence), the last terms in Eqs.~\ref{eq:Ampere_mod-full},~\ref{eq:Faraday_mod-full}, and ~\ref{eq:position_evol} vanish.

For boundary conditions, we impose periodic conditions along the $x$-axis. The vertical $z$-axis uses open boundaries, allowing particles and fields to exit the domain and preventing artificial accumulation of magnetic flux or excessive heating, as explained in detail in \citetalias{2024MNRAS.530.1866S}.

\subsection{Notation convention}

Here we introduce different kinds of averages denoted by angled bracket and different subscripts, namely $\langle A \rangle_x$, $\langle A \rangle_v$, $\langle A \rangle_g$, $\langle A \rangle_p$, and $\langle A \rangle_{nt}$. $\langle A \rangle_x\equiv\frac{1}{L_x}\int Adx$ denotes an average taken over the $x$ coordinate, the result will depend on the vertical coordinate $z$; the volume-average $\langle A \rangle_v\equiv\frac{1}{2HL_x}\int Adxdz$ is the average over the $2D$ volume delimited by the disk height $H$, where $H$ is calculated as ${H=(2k_B\overline{T}/m\Omega_0^2)^{1/2}}$, where $\overline{T}$ is a measure of the instantaneous temperature in the disk. Here we are also introducing the overline ($\overline{\phantom{X}}$) notation, that correspond to the ratio between two volume-averages, in particular we use this average to compute the average plasma magnetization in the disk ${\overline{\sigma}\equiv\langle B^2\rangle_v/\langle 4\pi nmc^2\rangle_v}$ and the disk average temperature $k_B\overline{T}=\langle P\rangle_v/\langle n \rangle_v$, where $P$ and $n$ are the isotropic pressure and particle density. We also define the global particle average ${\langle A \rangle_g\equiv\frac{1}{N}\sum_j A_j}$ is the particle-average taken from the particles that are within the disk, the particle average per cell ${\langle A \rangle_p\equiv\frac{1}{N_j}\sum_j A_j}$ is the average from particle within the same simulation cell ``j'', where $N_j$ is the number of particles in that cell; and the corresponding non-thermal particle average ${\langle A \rangle_{nt}\equiv\frac{1}{N_j}\sum_{j,\gamma_j> \gamma_c} A_{j}}$ is the particle average, considering only particles with final Lorentz factor larger than certain cut-off Lorentz factor (see below). All the previous averages results in time dependent quantities.

%In addition, we also introduce the overline ($\overline{\phantom{X}}$) notation, that correspond to the ratio between two volume-averages, in particular we use this average to compute the average plasma magnetization in the disk ${\overline{\sigma}\equiv\langle B^2\rangle_v/\langle 4\pi nmc^2\rangle_v}$ and the disk average temperature $k_B\overline{T}=\langle P\rangle_v/\langle n \rangle_v$, where $P$ and $n$ are the isotropic pressure and particle density.

\section{Turbulence development} \label{sec:turbulence}

\begin{figure*}%[ht!]
   \centering
   \includegraphics[width=\textwidth]{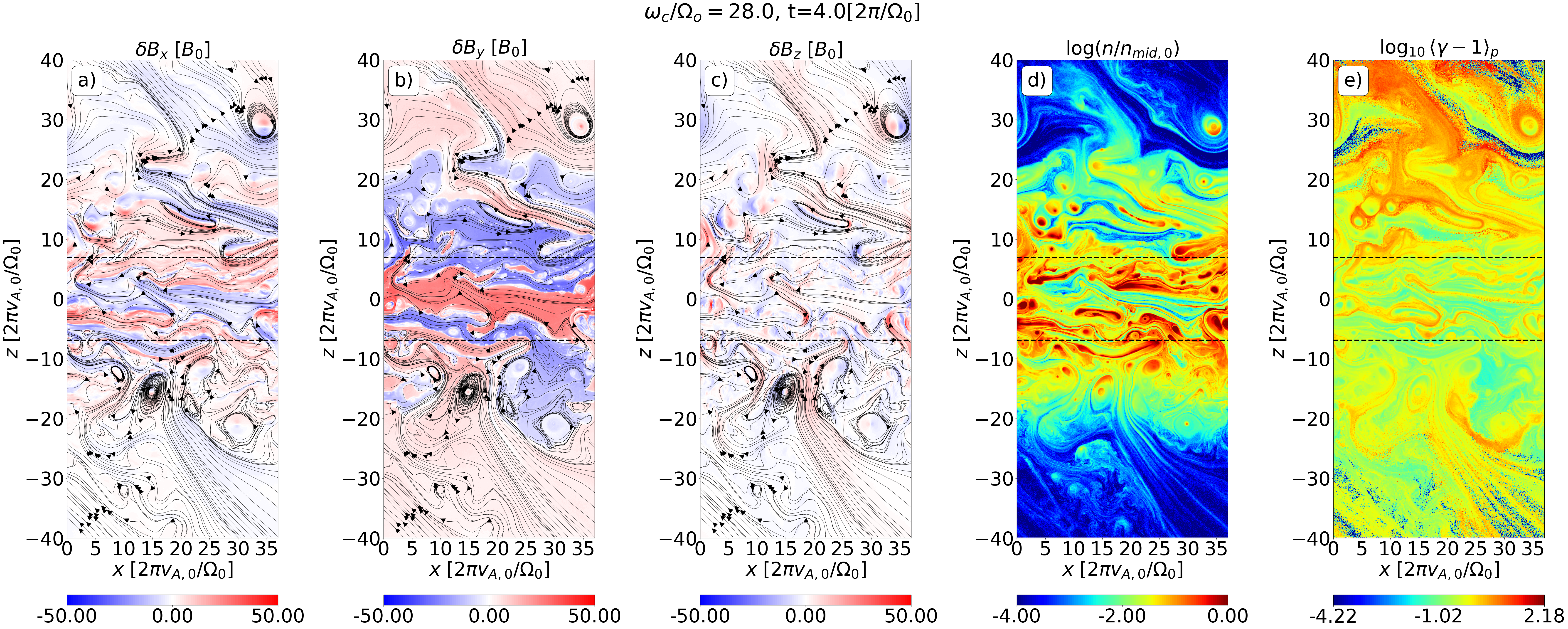}
   \caption{Components of the magnetic field fluctuations $\delta \mathbf{B} = \mathbf{B} - \mathbf{B}_{0,j}$, with arrows showing the poloidal field direction (panels $a$–$c$), plasma density $n$ (panel $d$), and the average particle kinetic energy $\langle \gamma - 1 \rangle_p$ per cell (panel $e$), for simulation ST2D-28 ($\omega_{c,0}/\Omega_0=28$) at $t=4\,[2\pi/\Omega_0]$. The horizontal dotted lines in all panels mark the disk scale height.}
   \label{fig:turbulence}%
\end{figure*}

We summarize the main features of stratified kinetic MRI turbulence in 2D, focusing on the properties relevant to particle acceleration. A more detailed description of the MRI evolution in this setup is provided in \citetalias{2024MNRAS.530.1866S}.

Figure~\ref{fig:turbulence} shows the magnetic field fluctuations ${\delta B_j = B_j - B_{0,j}}$ for the simulation with scale separation ${\omega_{c,0}/\Omega_0=28}$, where ``j'' denotes the cartesian component of the field, at $t=4\,[2\pi/\Omega_0]$: panels $a$–$c$ show the $x$, $y$, and $z$ components with streamlines indicating the poloidal field direction; panel $d$ shows plasma density; and panel $e$ shows the particle-average kinetic energy $\langle \gamma - 1 \rangle_p$. At this point, turbulence is fully-developed.

A large scale magnetic dynamo emerges during the turbulent phase, consistent with previous stratified MHD studies \citep{2013ApJ...767...30B,2016MNRAS.457..857S}, and evident in panels $a$ and $b$: the $x$ and $y$ components exhibit extended radial patches of single polarity with minor fluctuations, varying primarily along $z$. The dynamo is most prominent in the azimuthal $y$ component. Additionally, panels $a$-$c$ also reveal the presence of magnetic loops coincident with dense plasmoids (more clearly seen in $d$), produced by prior magnetic reconnection events, as explained in \citetalias{2024MNRAS.530.1866S}. %The plasma density peaks in regions of weak magnetic field. 

Plasmoids formed in the turbulence-driven magnetic reconnection are inefficient at trapping high-energy particles. This is seen by comparing Panels $d$ and $e$ of Figure \ref{fig:turbulence}, which show an anti-correlation between plasma density and the average energy of the particles. Unlike previous 2D reconnection studies \citep{2014ApJ...783L..21S}, where plasmoids trap high-energy particles slowing down their further acceleration \citep{2018MNRAS.481.5687P,2021ApJ...922..261Z}, plasmoids here are filled mainly with low-energy particles, while high-energy particles concentrate near the outskirts of high-density regions. Below we will see that this feature is directly related with the initial injection of the accelerated particles, which requires them to escape from splitting plasmoids.

Our simulations generate within the disk a high-energy particle population whose energy distribution is well-fitted by a power law with an exponential cutoff, as shown in \citetalias{2024MNRAS.530.1866S},
\begin{equation}
    \frac{dn}{d\gamma}\propto (\gamma-1)^{-p}e^{-\gamma/\gamma_c}, \label{eq:En_dist}
\end{equation}
which are characterized by the slope $p$ and the cutoff Lorentz factor $\gamma_c$.

\begin{figure}%[ht!]
    \centering
    \includegraphics[width=1.0\linewidth]{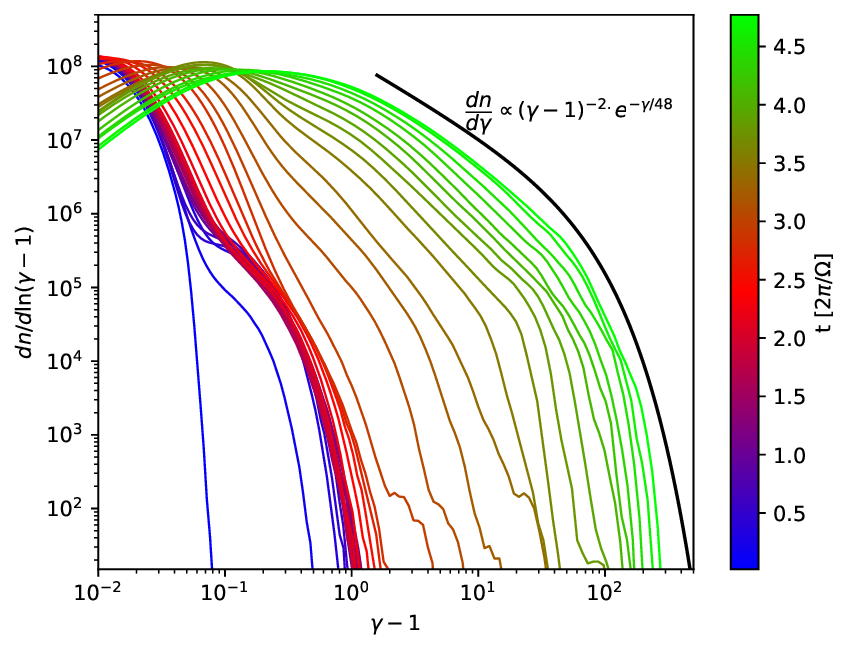}
    \caption{Time evolution of the particle energy spectrum within the disk ($|z| < H$) in simulation ST2D-28. The simulation runs until $t=5\,[2\pi/\Omega_0]$, period over which the plasma temperature increases from $k_B\overline{T}=5\times10^{-3}$ to $3\times10^{-1}\,mc^2$. The black line shows the power-law fit with exponential cutoff for the final distribution, with slope $p\simeq2$ and cutoff Lorentz factor $\gamma_c\simeq48$.}
    \label{fig:spect28}
\end{figure}

Figure~\ref{fig:spect28} shows the evolution of the particle energy spectrum for run ST2D-28, considering only the particles within the disk ($|z| < H$). Acceleration rate occur fastest between $t\approx 2$ and $3\,[2\pi/\Omega_0]$, as can be seen from the transition of a nearly thermal particle energy distribution to a non-thermal distribution with a high energy component, As explained in \citetalias{2024MNRAS.530.1866S}, this coincides with the initial channel-flow disruption due to magnetic reconnection. Acceleration persists during the subsequent turbulent phase shown in Fig.~\ref{fig:turbulence} ($t=4[2\pi/\Omega_0]$), consistent with \citet{2024PhRvL.133d5202B}. After $t=3\,[2\pi/\Omega_0]$, the high-energy tail hardens and extends to larger $\gamma$.

By the end of the simulation (${t\approx 5\,[2\pi/\Omega_0]}$, when ${k_B\overline{T}\approx0.3\,mc^2}$), the spectrum is well described by a power-law with exponential cutoff with spectral index $p\simeq2$ and cutoff Lorentz factor $\gamma_c\simeq48$, shown by the black line in Fig. \ref{fig:spect28}. This cutoff grows with $\omega_{c,0}/\Omega_0$ as shown in \citetalias{2024MNRAS.530.1866S}, which is consistent with acceleration being caused purely by magnetic reconnection, in which the maximum attainable energy is limited to particle having Larmor radius of the size of the reconnecting region \citep[][see \citetalias{2024MNRAS.530.1866S} for more details]{2016ApJ...816L...8W}.

\section{Injection and acceleration of high-energy particles} \label{sec:acc28}

We analyze the injection process and the subsequent post-injection acceleration of high-energy particles in run ST2D-28, focusing on identifying the underlying mechanisms and their localization within the disk. To this end, we track a subset of $N_{track} = 491$ high energy particles that reach energies larger than the cutoff energy and remain within the disk region (defined by $|z| < H$) by the end of the simulation (when $k_B\overline{T}=0.3\,mc^2$).

The other parameters of tracked particles in run ST2D-28 are in Table~\ref{tab:cut_off}.
%Table \ref{tab:cut_off} summarizes key parameters: the cut-off Lorentz factor $\gamma_c$ at the end of each simulation, used to select the tracked particles that have a Lorentz factor larger than $\gamma_c$; number of tracked particles $N_{\text{track}}$; the Lorentz factor at injection $\gamma_{\text{inj}}$ (defined below); the tracking start time $t_{\text{track}}$; and sampling resolution $\delta t_{track}$.

In the shearing frame, the particle energization rate follows:
\begin{equation}
mc^2\frac{d\gamma}{dt}=q\mathbf{E}\cdot\mathbf{v}+\frac{3}{2}\Omega_0 p_x v_y - m\Omega_0^2 z v_z, \label{eq:Energy_evol}
\end{equation}
where the first term represents work by the electric field, the second term corresponds to global shear energization \citep{2012ApJ...755...50R}, and the last term is the work done by the vertical component of gravity.\footnote{This is the particle energy perceived by a local observer that moves with the shear velocity of the disk.} Since we focus on energetic particles that remain in the disk, their vertical displacement is minimum, so the contribution of the vertical component of gravity to their energization is negligible.

For particles reaching Lorentz factors $\gamma \geq \gamma_c = 48$ by the end of the simulation, we track trajectories, velocities, and local fields for ${t \geq t_{track} = 2\,[2\pi/\Omega_0]}$.

\begin{figure}%[ht!]
\centering
\includegraphics[width=\linewidth]{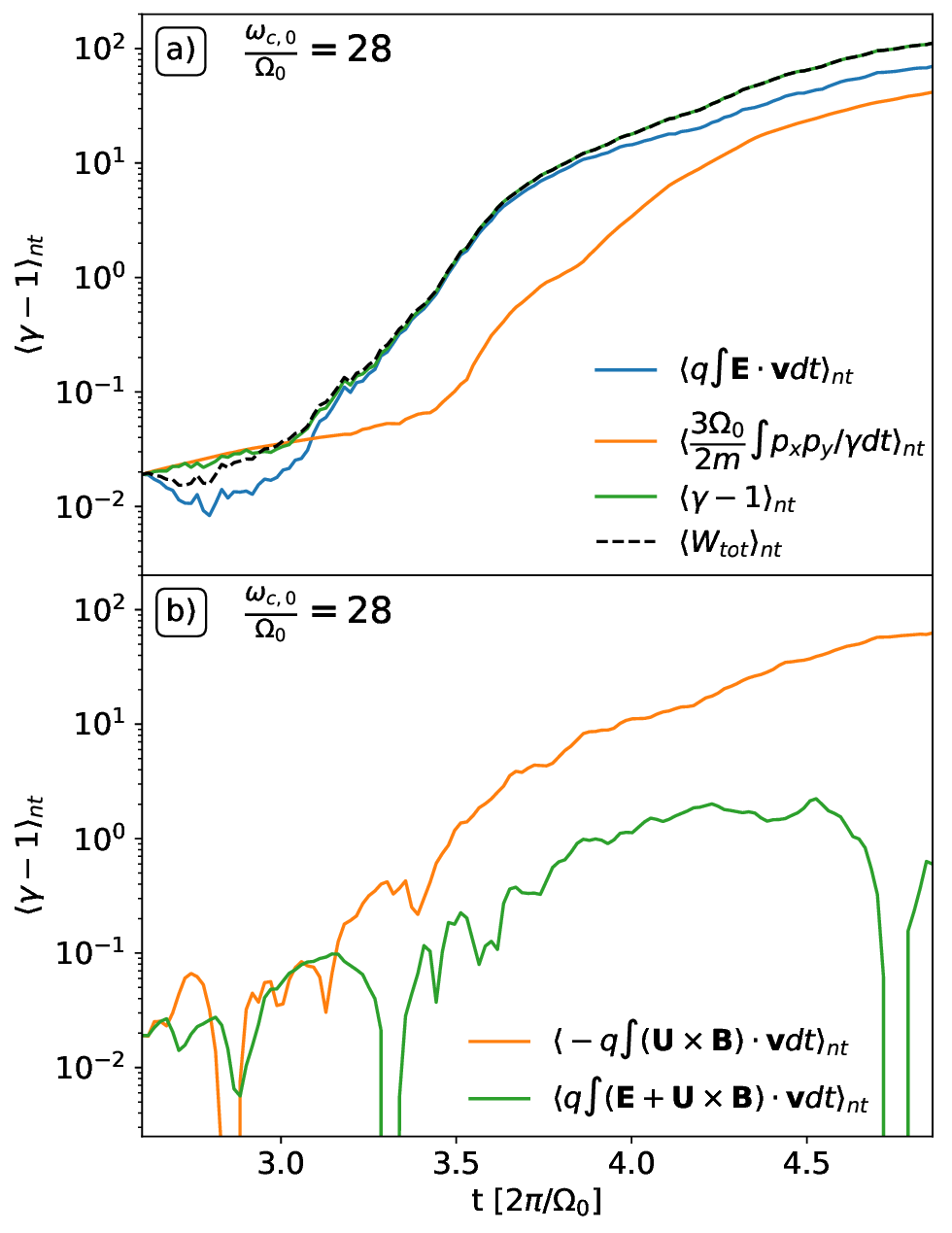}
\caption{Panel $a$: Time evolution of the average kinetic energy $\langle \gamma-1 \rangle_{nt}$ (green) of tracked particles in ST2D-28. The blue and orange lines show the energy gain from the electric field and shear, respectively, and the black dashed line shows their sum. Panel $b$: Decomposition of the work done by the ideal (orange) and non-ideal (green) electric field.  } %The starting time is chosen to avoid noise in the energy integration that occurs at low particle energies.}
\label{fig:meanE_simt}
\end{figure}

Figure \ref{fig:meanE_simt}a  shows the evolution of average kinetic energy $\langle \gamma - 1 \rangle_{nt}$ (green) from $t=2.6$ [$2\pi/\Omega_0$].\footnote{The starting time is chosen to avoid noise in the energy integration that occurs at low particle energies.} Blue and orange curves show contributions from electric field work and shear, respectively; their sum (black dashed) matches the energy averaged over all the energetic particles, ${\langle\gamma - 1\rangle_{nt}}$. During the final part of the simulation (${4.3\leq t\Omega_0/2\pi<4.8}$), approximately two-thirds of the energization originates from electric work, with the remaining third from shear.
Figure \ref{fig:meanE_simt}b decomposes the average electric field work into ideal ${\langle -q\int(\frac{\mathbf{U}}{c}\times\mathbf{B})\cdot \mathbf{v}dt\rangle_{nt}}$ (where $\mathbf{U}\equiv \langle \mathbf{v}\rangle_p$ is the plasma bulk velocity) and non-ideal components ${\langle q\int(\mathbf{E}+\frac{\mathbf{U}}{c}\times\mathbf{B})\cdot \mathbf{v}dt\rangle_{nt}}$, revealing that nearly all electric work arises from the ideal field. However, we prove below that the non-ideal fields play a crucial role in the injection of particles into their initial non-thermal acceleration, which is driven by the reconnection of the magnetic fields of ``splitting'' plasmoids. This is followed by a second stage, in which particles are accelerated by a fast first-order Fermi-like process. Finally, the acceleration is completed by a slower, second-order Fermi-like acceleration, complemented by a subdominant contribution from shear acceleration.

\begin{deluxetable*}{ccccc}
\tablewidth{0pt}
\tablecaption{Sampled particle parameters \label{tab:cut_off}}
\tablehead{
\colhead{Run} & \colhead{ST2D-28} & \colhead{ST2D-14} & \colhead{ST2D-7} & \colhead{ST2D-3.5} 
}
\startdata
$\omega_{c,0}/\Omega_0$ & 28 & 14 & 7 & 3.5 \\
$\gamma_c$ & 48 & 25 & 12 & 5 \\
$N_{track}$ & 491 & 265 & 219 & 251 \\
$\gamma_{inj}$ & 1.5 & 1.2 & 1.5 & 1.6 \\
$t_{track}$ [$2\pi/\Omega_0$] & 2 & 2 & 2 & 1.5 \\
$\delta t_{track}$ [$\Delta/c$] &20.25 & 20.25 &20.25 &10.125\\
\enddata
\tablecomments{ Parameters of the sampled high-energy particles: final cut-off Lorentz factor $\gamma_c$, number of tracked particles $N_{track}$, Lorentz factors at injection $\gamma_{inj}$, the starting tracking time $t_{track}$, and the time resolution of the tracking particles. }
\end{deluxetable*}

%\subsection{Initial reconnection-driven acceleration} \label{sec:inj}
\subsection{The first two stages: reconnection-driven injection and first-order Fermi} \label{sec:inj}

\begin{figure*}%[ht!]
\centering
\includegraphics[width=\textwidth]{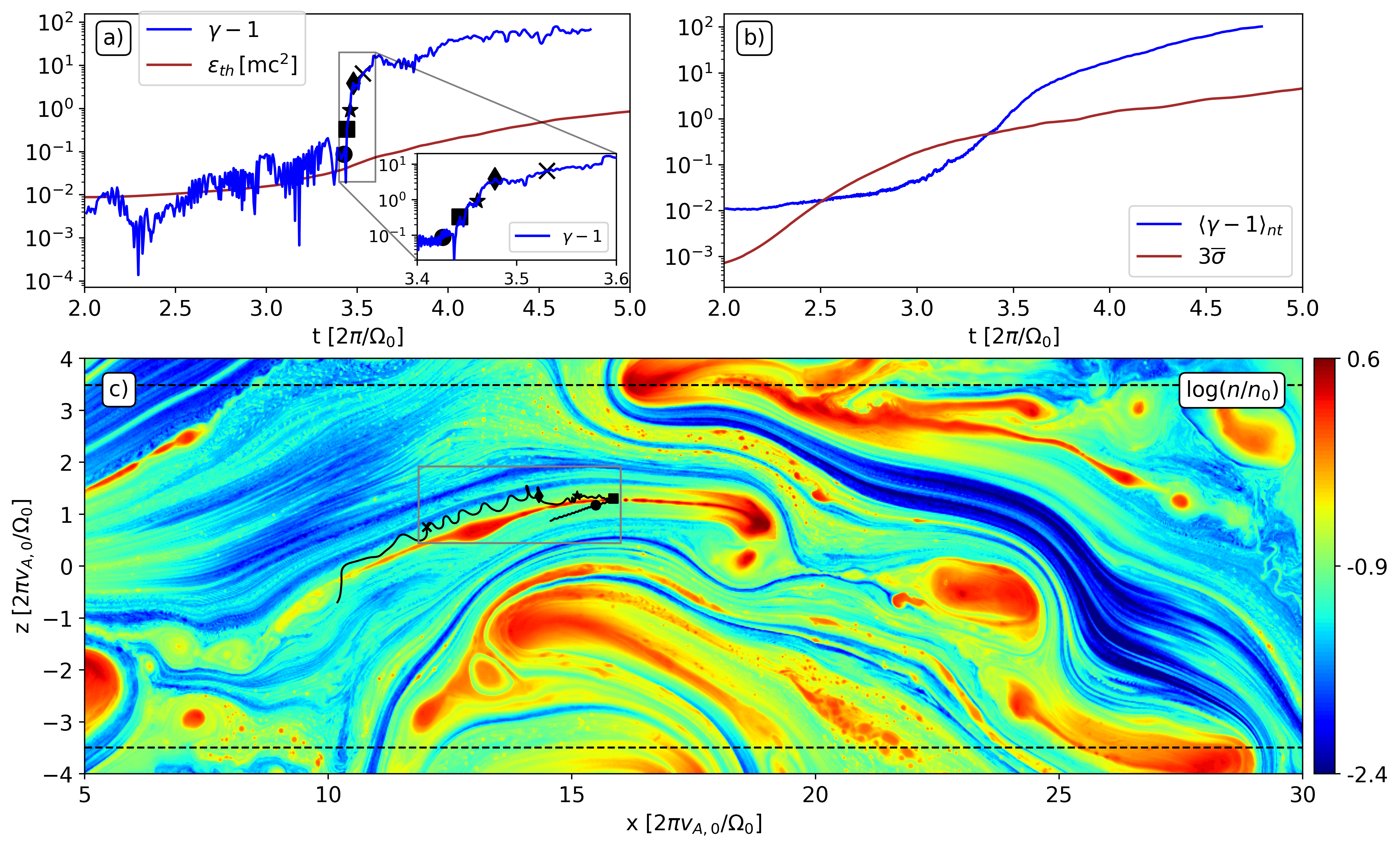}
\caption{Particle injection process in run ST2D-28. Panel~$a$: Evolution of a representative particle’s energy (blue) and plasma average kinetic energy $\varepsilon_{th}$ (brown). The black circle, square, star, diamond and cross mark the different stages of the injection. Panel~$b$: Average non-thermal particle energy, $\langle \gamma -1\rangle_{nt} $ (blue) and $3\overline{\sigma}$ (brown). Panel~$c$: Particle trajectory over plasma density; dashed lines delimit the disk scale height.}% Panels~$d$–$f$: Zoomed snapshots showing the particle (shaped marks) and its trajectory (black line) over plasma density and magnetic field streamlines at the beginning, middle, and end of the injection process. Panel~$g$ same as Panel~$e$ but over $B_y$, showing that injection occurs near $B_y \approx 0$.}
\label{fig:trajectory}
\end{figure*}

The acceleration of the energetic particles in our runs is characterized by a one-shot initial jump in their energy, which we identify as their ``injection''. This is illustrated in Figure~\ref{fig:trajectory} which shows the acceleration process for a representative particle. Panel~$a$ shows the particle’s energy in blue alongside the average thermal energy of all particles within the disk ${\varepsilon_{\text{th}}\equiv \langle \gamma-1 \rangle_g mc^2}$ in brown. Around $t \sim 3.5$ [$2\pi/\Omega_0$], the particle undergoes a rapid energy increase, exceeding ${\varepsilon_\text{th}}$ by a factor $\sim 100$.

In order to prove the importance of non-ideal fields in this injection process, we recalculate the average  works done by the ideal and non-ideal electric fields as in Figure~\ref{fig:meanE_simt}$b$, but as a function of the post-injection time of each particle, $t^*$. This way we can clearly see what is the dominant average contribution at injection ($t^*=0$). For this, we need to device a general injection criterion applicable to all the particles. We do this by following an analogous approach as \citet{2021ApJ...922..261Z,2023ApJ...956L..36Z}, who use $\gamma \geq 3\sigma$ as an injection criterion for acceleration due to relativistic reconnection.  Since in our case we expect injection to happen in an evolving, highly inhomogeneous environment, we first determine an injection energy as the one that occurs when ${\gamma_{\text{inj}}-1=\langle \gamma -1 \rangle_{nt} = 3 \overline{\sigma}}$, where ${\overline{\sigma} \equiv \langle B^2\rangle_v/\langle 4\pi nmc^2\rangle_v}$. Panel b of Figure~\ref{fig:trajectory} shows the average particle energy in blue, with $3\overline{\sigma}$ in brown. 
%Following \citet{2021ApJ...922..261Z,2023ApJ...956L..36Z}, we first estimate the average energy attained by the particles at the moment of their injection. We do that by defining first a ``global'' injection time as the time when $\langle \gamma - 1 \rangle_{nt} = 3\sigma$. 
This allows us to find the injection energy ${\gamma_\text{inj}-1 \approx 0.5}$, which occurs at ${t\sim3.4\, [2\pi/\Omega_0]}$. 
Thus, using the criterion ${\gamma_{\text{inj}} - 1 \approx 0.5}$, we define each particle’s injection time $t_{\text{inj}}$, which allows us to re-express their evolution in terms of the post-injection time $t^* = t - t_{\text{inj}}$.\footnote{Because particles are injected at different simulation times, aligning them by $t_\text{inj}$ causes the extension of their post-injection times to be different. To minimize bias, we stop our post-injection analysis once $15\%$ of the tracked particles have exited the dataset.} 

\begin{figure}%[ht!]
\centering
\includegraphics[width=1\linewidth]{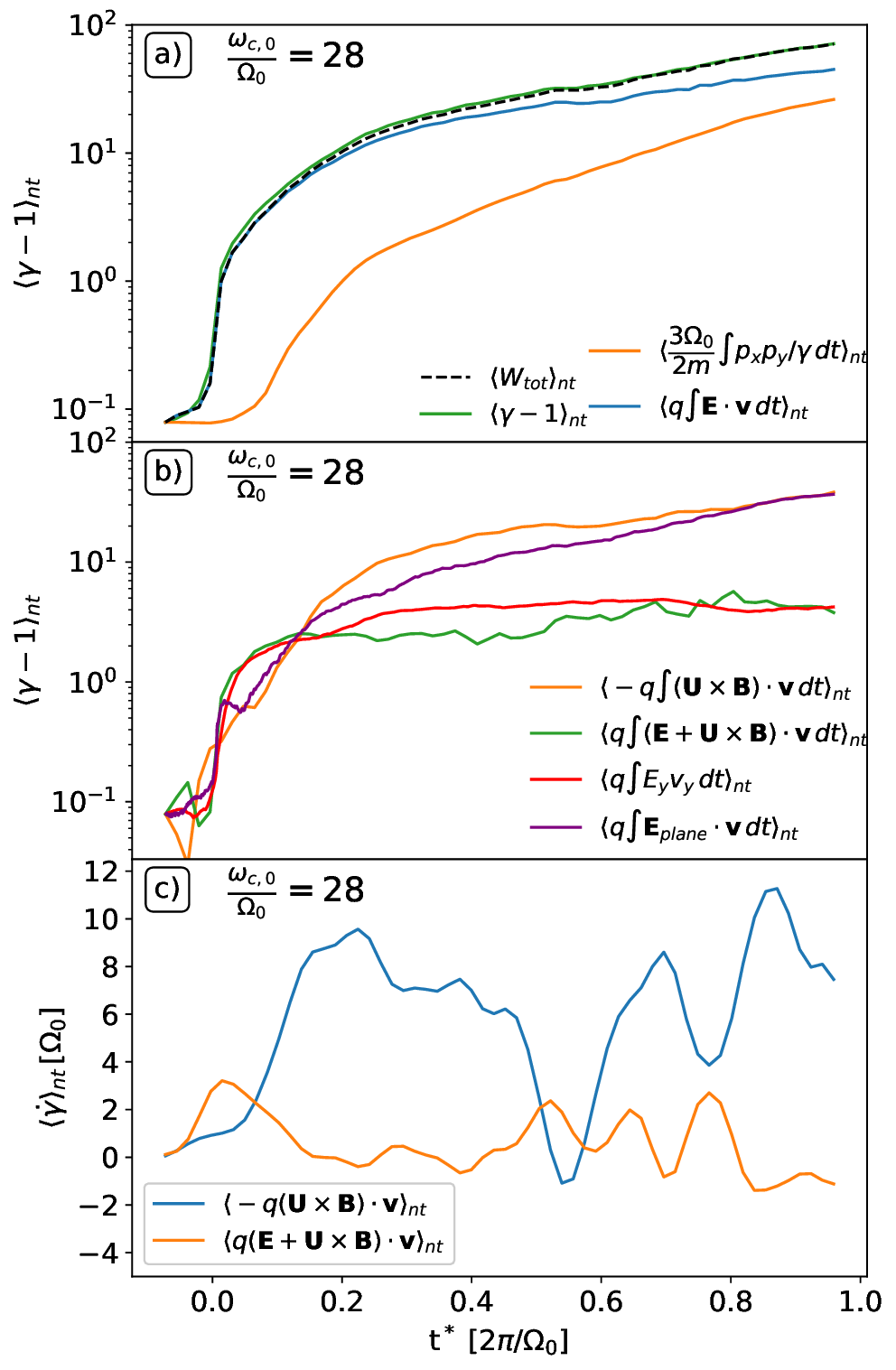}
\caption{Panels $a$ and $b$ show the same as Fig.~\ref{fig:meanE_simt}, but as a function of injection-aligned time $t^* = t - t_{\text{inj}}$. In addition, panel $b$ shows the decomposition of the work done by the azimuthal (red) and poloidal (purple) electric field.  The average is computed over the first 354 injected particles. Panel $c$ shows the power done by the ideal (blue) and non-ideal (orange) electric fields.}
\label{fig:injection_28}
\end{figure}

Figure~\ref{fig:injection_28}$a$ shows the same quantities as Fig.~\ref{fig:meanE_simt}$a$, but now as a function of $t^*$. 
The most important change is that there is clearly a very rapid increase in energy at $t*\approx 0$, which characterizes the one-shot injection of particles and is almost exclusively driven by the electric field. Shear acceleration eventually reaches about 1/3 of the total energization as also shown in Fig. \ref{fig:meanE_simt}$a$, but this occurs only at the end. Further characterization of the injection process is obtained from figure \ref{fig:injection_28}$b$, which shows the same quantities as Fig. \ref{fig:meanE_simt}$b$, but in terms of $t^*$. We can see that the non-ideal electric field drives the initial energy rise to $\langle\gamma-1\rangle_{nt} \sim 1$ during injection (${0\leq t^*\leq0.067\,[2\pi/\Omega_0]}$) but becomes subdominant afterward. This can be seen more clearly from Fig. \ref{fig:injection_28}$c$, which shows the time derivative of the energy contributions from the ideal and non-ideal electric fields.\footnote{In order to calculate these derivatives we smooth the energy contributions from the ideal and non-ideal fields using a Gaussian filter with window ${0.026\, [2\pi/\Omega_0]}$} The rapid non-ideal energization during injection coincides with the work done by the azimuthal component of the electric field (red line), corresponding to the electric field responsible for the reconnection of the poloidal (in plane) components of the magnetic field.
Subsequently we enter the second stage, where the energy gained is dominated by the ideal electric field, whose direction is mainly poloidal, as shown by the purple line. This behavior is thus consistent with a two-stage initial acceleration where an initial non-ideal reconnection-driven injection is characterized by particles drifting in the azimuthal direction, while a subsequent energy gain dominated by an ideal electric field is caused by particles drifting in the poloidal direction.

We further explore this initial two-stage acceleration process by analyzing the trajectory and energy evolution of a typical accelerated particle. Figure~\ref{fig:trajectory}$c$ shows the orbit of a non-thermal particle during the different stages of its acceleration process, overlaid on the plasma density, with dashed lines denoting $|z|=H$. The black circle, square, star, diamond, and cross symbols mark the position of the particle at the corresponding times denoted by the same symbols in panel $a$. The time corresponding to the black square (at the beginning of the injection process) coincides with the time of the overlaid density.

\begin{figure*}%[ht!]
\centering
\includegraphics[width=\textwidth]{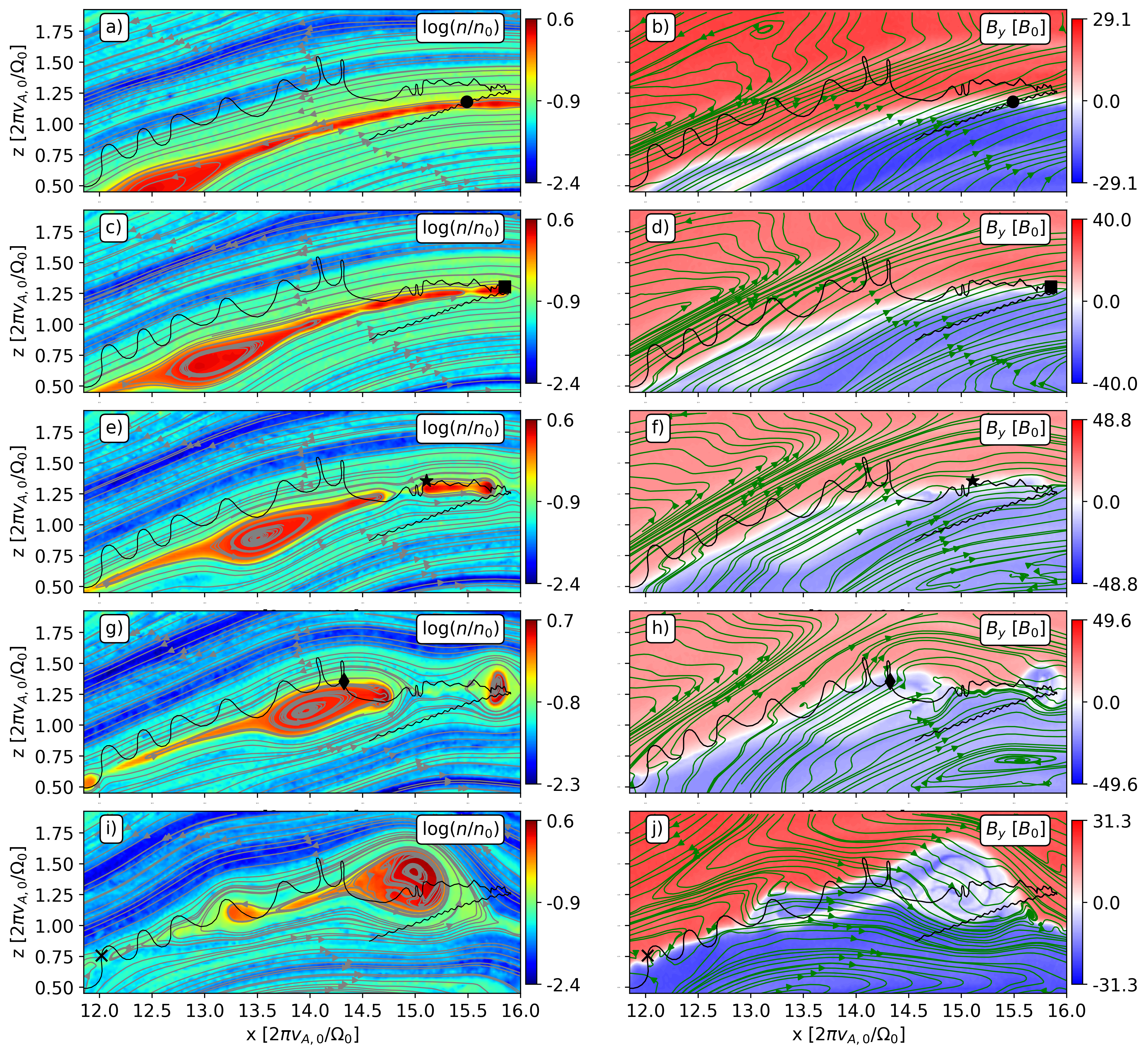}
\caption{Zoomed snapshots of the particle showed in Figure \ref{fig:trajectory}, the corresponding particle position are shown by the shaped marks over the plasma density and magnetic field streamlines in gray (left column); and over the azimuthal component of the magnetic field $B_y$ and plasma velocity streamlines in green (right column). Right columns show that injection occurs near $B_y \approx 0$.}
\label{fig:trajectory_zoom}
\end{figure*}

Figure \ref{fig:trajectory_zoom} shows a zoom in on the gray rectangle of Fig.~\ref{fig:trajectory}$c$. Left (right) panels show a zoom-in on the background density (azimuthal magnetic field $B_y$) at the times corresponding to the different particle's positions marked by the black circle, square, star, diamond, and cross symbols in Figure ~\ref{fig:trajectory}$c$. In these figures, the particle’s trajectory shown in figure ~\ref{fig:trajectory}$c$ is plotted as a black line, and the gray (green) streamlines in the left (right) panels show the direction of the local magnetic field (plasma velocity) in the $x-z$ plane. Initially (panels $a$ and $b$), the particle resides in an elongated high-density plasmoid that separates regions of poloidal and azimuthal magnetic fields of opposite polarity, as can be seen from the gray lines and from the background colors in panels $a$ and $b$, respectively. Reconnection of the poloidal component of the magnetic field disrupts this plasmoids, which splits into two newly formed plasmoids that move away from each other (panels $c$ and $d$), allowing the particle to move around the corresponding reconnection X-point that forms between them, while it gets accelerated by the azimuthal non-ideal electric field. This stage gives rise to the first rapid increase in the particle energy, which can be seen from Fig. \ref{fig:trajectory}$a$ by comparing the particle's energy at the times of the square and circle symbols. After this injection stage, the particle starts drifting in a Speiser-like orbit along the large-scale current sheet formed by the mainly azimuthal  magnetic field \citep{1965JGR....70.4219S}. This can be seen by looking at the part of the trajectory marked by the star symbol in Fig.~\ref{fig:trajectory}$c$ and more clearly in Figs.~\ref{fig:trajectory_zoom}$e$ and ~\ref{fig:trajectory_zoom}$f$. During this stage, the particle continues to undergo very rapid energy increase, as can be seen from Fig.~\ref{fig:trajectory}$a$. The green lines in Figure~\ref{fig:trajectory_zoom}$f$ show that at the particle's position there is a vertical convergence of plasma velocity associated to the plasmoid splitting dynamics (star symbol in Fig.~\ref{fig:trajectory_zoom}$f$). This suggests that the energy gain during this Speiser-like regime comes from the ideal electric field associated to converging plasma motions, giving rise to a first-order Fermi-like process, consistent with the ideal energy gain shown on average by the particles in Fig~\ref{fig:injection_28}$c$.

This second acceleration stage, characterized by a first-order Fermi-like acceleration, essentially finishes right before the time marked by the diamond symbol in Fig.~\ref{fig:trajectory}$a$. We can see from Figs.~\ref{fig:trajectory_zoom}$g$ and ~\ref{fig:trajectory_zoom}$h$ that during this period the particle moves inside a plasmoid. Finally (panels $i$ and $j$), the particle continues to move along the current sheet formed by the large-scale azimuthal and poloidal magnetic fields, gaining energy at a highly time-variable rate but achieving a slow energy increase overall. This behavior is reminiscent of an stochastic energy gain taking over in the final stage of the acceleration process. In Section \ref{sec:post_acc28} we see that this is on average the case when considering the entire non-thermal particle population.

\begin{figure}%[ht!]
\centering
\includegraphics[width=1\linewidth]{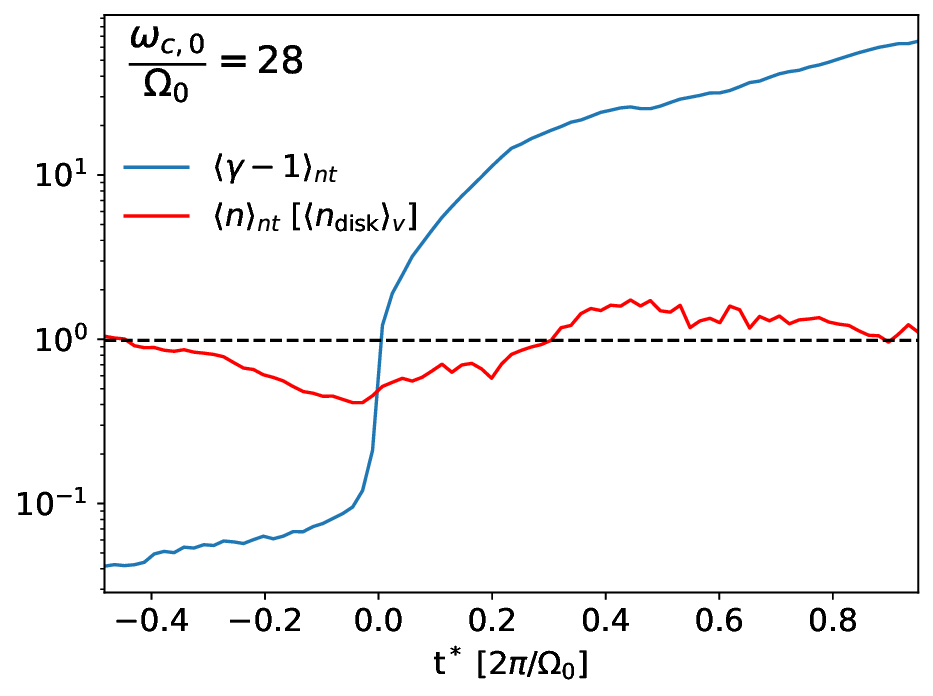}
\caption{Average particle energy (blue) and local background density (red) at the particle positions, normalized by the disk-averaged density, as a function of injection-aligned time $t^*$. The horizontal dashed line shows the density averaged between $t^* = -0.5$ and $0.9$ [$2\pi/\Omega_0$].}
\label{fig:density}
\end{figure}

Panels $a$ and $c$ of Figure \ref{fig:trajectory_zoom} show that the injected particles escape from a high- to low-density region. Figure~\ref{fig:density} illustrates that this is a systematic behavior during the injection by showing $\langle \gamma-1\rangle_{nt}$ (blue) and the mean background density at particle positions $\langle n\rangle_{nt}$, normalized by the instantaneous disk-averaged density $\langle n_{\text{disk}}\rangle_v$ (red). We can see that a  minimum in $\langle n\rangle_{nt}$ occurs right around the time of injection, indicating the escape from dense regions suggested by our 1-particle analysis shown in Fig. \ref{fig:trajectory_zoom}. The pre-injection drop of $\langle n\rangle_{nt}$ can be explained %if, before injection, particles are moving near the edge of the high density region as can be seen in panels $d$-$f$ of Figure \ref{fig:trajectory} decreasing the density previous to the injection. Furthermore,
given that our definition of injection energy leads to a dispersion in the actual injection times of particles, which contributes to the early decrease in the average density. The subsequent rise corresponds to interactions with high-density plasmoids as the particles continue drifting along the current sheet formed by the large-scale magnetic field, which we showed in panels $i$ and $j$ of Figure \ref{fig:trajectory_zoom} to be the case for our analyzed representative particle. This behavior is also consistent with panel $e$ of Figure \ref{fig:turbulence} showing that the high energy particles distribute near, but out of, the high-density plasmoids. As the thermal particles group in regions with low magnetic pressure or inside plasmoids, high energy particles have enough energy to move toward more magnetized regions or escape from plasmoids.

\subsection{Second-order Fermi-dominated acceleration} \label{sec:post_acc28}

\begin{figure}%[ht!]
\centering
\includegraphics[width=1\linewidth]{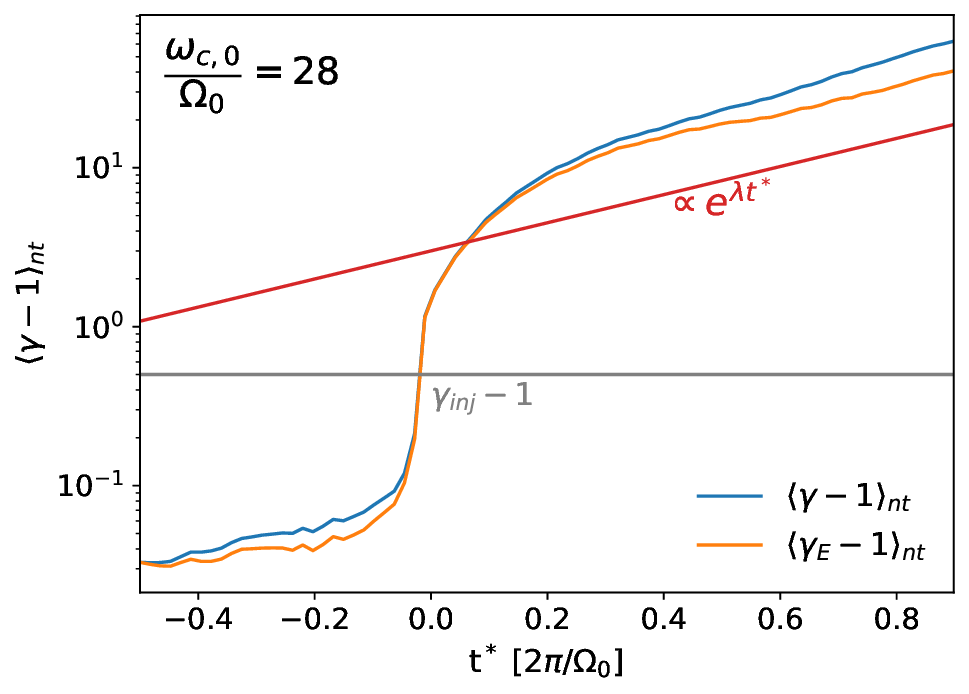}
\caption{Evolution of the average kinetic energy $\langle \gamma-1\rangle_{nt}$ (blue) and electric contribution $\langle \gamma_E-1\rangle_{nt}$ (orange) for the first $80\%$ of the injected particles in ST2D-28. The gray line marks the injection energy, and the red line shows the predicted second-order Fermi acceleration.}
\label{fig:Fermi28}
\end{figure}

The energy evolution followed by the particle analyzed in Figs. \ref{fig:trajectory} and \ref{fig:trajectory_zoom} suggests that after their reconnection-driven injection and first-order Fermi-like acceleration, particles keep accelerating at a lower rate. In Figure~\ref{fig:Fermi28} we show the average particle energy in blue, and the isolated contribution from the electric field, ${\langle \gamma_E-1\rangle_{nt}}$ in orange. We see that the evolution of ${\langle \gamma_E - 1 \rangle_{nt}}$ is in good agreement with exponential growth with rate $\lambda \approx0.3\, \Omega_0$.
This behavior is in agreement with predictions from stochastic (second-order Fermi) acceleration mechanism \citep{2005A&A...441..845D,2011ApJ...735..102K,2012PhRvL.108x1102K,2012SSRv..173..557L,2014ASPC..488....8D,2016MNRAS.463.4331D,2020ApJ...899..151K,2022JPlPh..88a9014U}. In this mechanism, particles gain energy by scattering off moving magnetic structures, therefore the evolution of their average energy $\langle \gamma-1\rangle_{nt}$  depends on the velocity of these structures and is given by
\begin{equation}
\frac{d}{dt}\langle \gamma-1\rangle_{nt} = \lambda \langle \gamma-1\rangle_{nt}, \label{eq:Fermi2}
\end{equation} 

\begin{figure}%[ht!]
\centering
\includegraphics[width=1.0\linewidth]{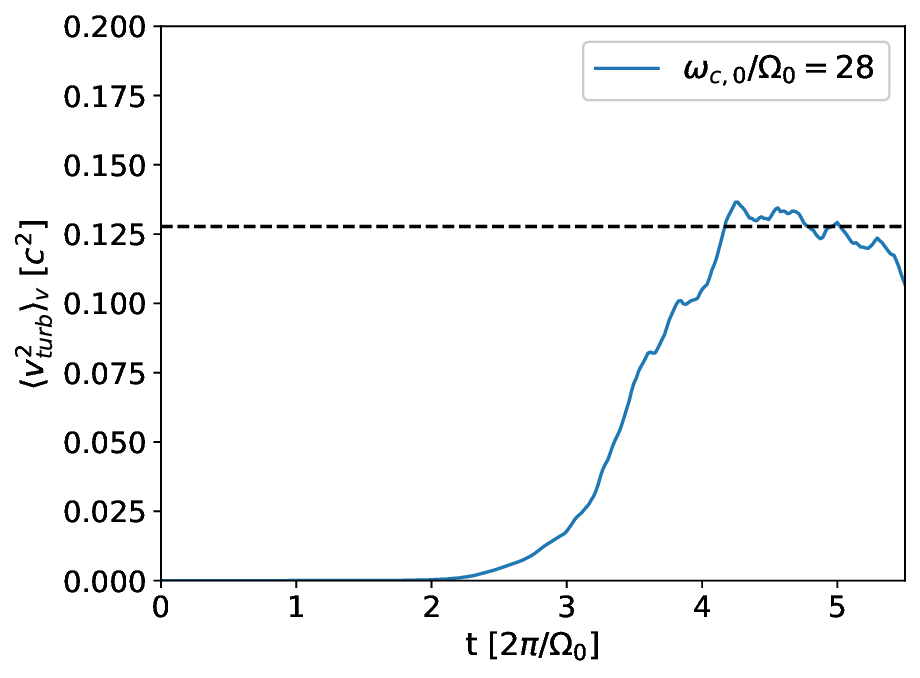}
\caption{Evolution of the squared fluid velocity averaged over the disk region for simulation ST2D-28. Dashed lines show average over $t=4$–$5$ [$2\pi/\Omega_0$].}
\label{fig:flow_vel}
\end{figure}

\noindent where the rate $\lambda \approx (v_{\text{turb}}/c)^2 c / \ell_{\text{mfp}}$, with $\ell_{\text{mfp}}$ being the average traveled distance by an accelerated particle between two consecutive scatterings \citep{2022JPlPh..88a9014U}. From \citetalias{2024MNRAS.530.1866S}, we estimated that ${\ell_{\text{mfp}} \approx (2\pi)^2 v_{A,0}/\Omega_0}$, independently from $\omega_{c,0}/\Omega_0$. 
Figure~\ref{fig:flow_vel} shows the evolution of the volume-averaged squared velocity $\langle v^2\rangle_{v}$ for the simulation with scale separation ${\omega_{c,0}/\Omega_0=28}$, with the horizontal dashed line marking the average between ${t=4-5\, [2\pi/\Omega_0]}$, when the post reconnection-driven acceleration dominates. We see that the average squared turbulent velocity is close to $\langle v_{turb}^2\rangle_t\approx 0.125c^2$. Thus, considering that in our runs $v_{A,0} = 10^{-2} c$, second-order Fermi acceleration should produce ${\lambda \approx 0.32\, \Omega_0}$, in good agreement with the growth rate of $\varepsilon$, ${\lambda \approx 0.3\, \Omega_0}$, obtained from our simulation.

\begin{figure}%[ht!]
\centering
\includegraphics[width=1\linewidth]{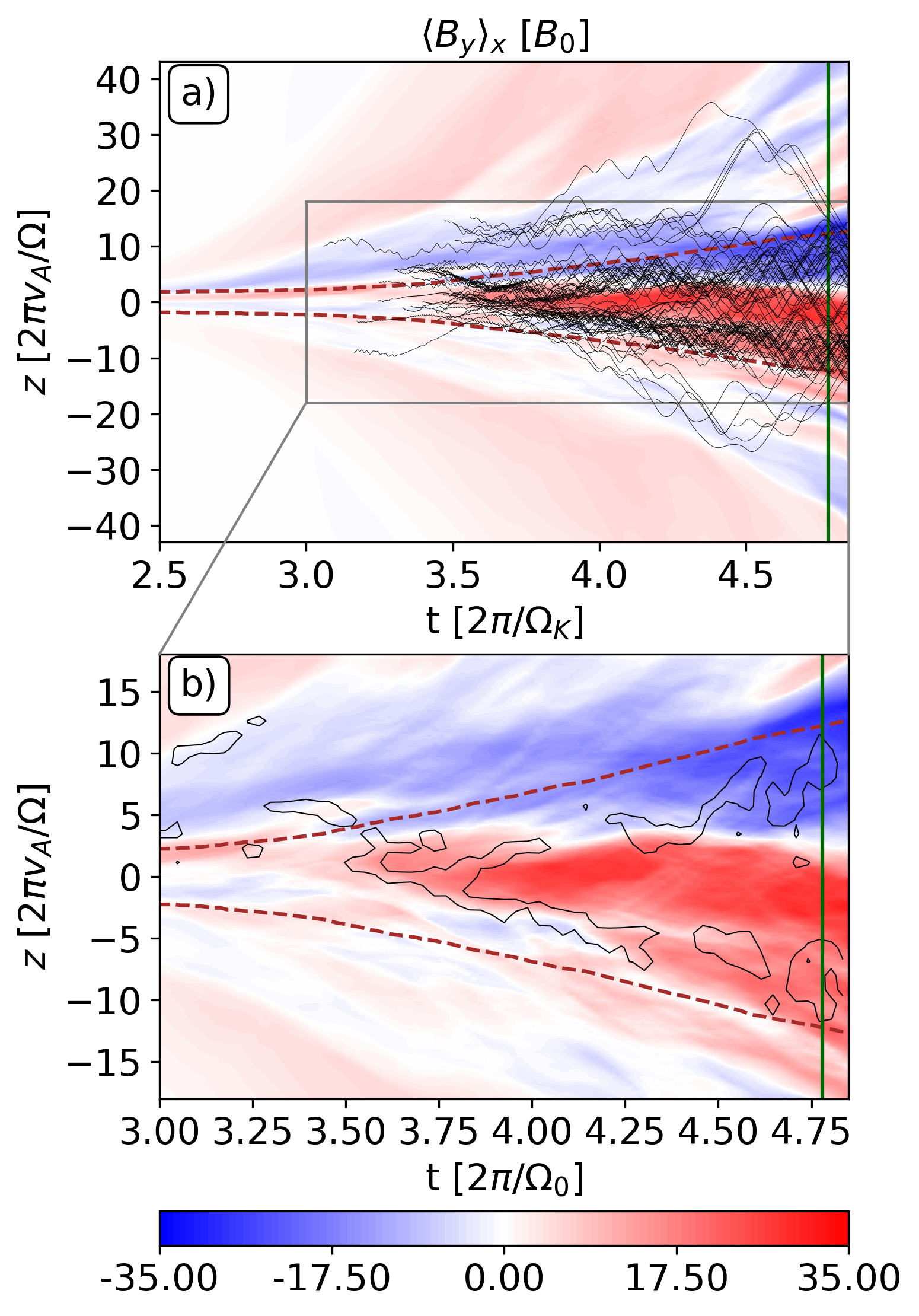}\\
\caption{Panel~$a$: Trajectories of 83 injected particles (black lines) over the dynamo diagram of $\langle B_y\rangle_x$ for ST2D-28. Dashed lines delimit the disk region. Panel~$b$: Particle density contour line in the portion of the height–time space diagram limited by the gray box of panel~$a$, showing particle accumulation near $B_y \approx 0$. The contour lines correspond to the region where the density drops to 75\% of its maximum value.}
\label{fig:dynamo_inj}
\end{figure}

Regarding the location within the disk where acceleration takes place, the trajectory followed by the particle analyzed in Figs.~\ref{fig:trajectory} and ~\ref{fig:trajectory_zoom} suggests that the large-scale dynamo field is important in confining the particles to regions of low azimuthal and poloidal large-scale magnetic fields as they get accelerated. This behavior suggests that the reconnection events that inject particles, along with the first-order Fermi-like stage and the final second-order Fermi plus shear acceleration dominated stage occur mainly in the low guide field regime. Here we show that this feature of the acceleration, is a general characteristic of the acceleration process, not limited to the 1-particle result obtained from Figs. \ref{fig:trajectory} and \ref{fig:trajectory_zoom}. Figure~\ref{fig:dynamo_inj}$a$ shows the trajectories of 83 high-energy particles (black dotted lines) since the moment of injection, plotted on top of  $\langle B_y\rangle_x(t,z)$. The particles tend to get concentrated near heights $z$ where $\langle B_y\rangle_x$ vanishes. Panel~$b$ shows the density contour of all tracked particles in the portion of the diagram limited by the gray box in Panel~$a$; the contour line marks the region where the density drops to a $75\%$ of its maximum value at each time. 
The particle density contours confirm their concentration along $\langle B_y\rangle_x \approx 0$ layers for times $t\lesssim 4.5\,[2\pi/\Omega_0]$. However, by the end of the simulation ($t\gtrsim 4.5\,[2\pi/\Omega_0]$) the density contours of high energy particles appear significantly less confined to $\langle B_y\rangle_x \sim 0$ layers, which is indicative that particles tend to leave the current sheet once they have reached their maximum energy.

In summary, the overall acceleration of particles appears to be a multi-stage process, initiated by a reconnection-driven non-ideal injection stage, characterized by the particles escaping from splitting plasmoids, followed by a second stage in which the particles gain energy through a first-order Fermi process. In this stage, the energy is provided by convective (ideal) electric fields, due to vertically converging flows that enter the region left by the separating plasmoids after they are formed by splitting. After this, there is a third stage, characterized by the combination of stochastic and shear acceleration. This final stage is consistent with previous test particles studies of particle acceleration in the MRI turbulence \citep{2021MNRAS.506.1128S}. The fact that this whole process is mainly confined to the current sheets produced by the large-scale azimuthal and poloidal magnetic fields underlines the importance of the MRI dynamo in the three stages of this acceleration process.

\section{$\omega_{c,0}/\Omega_0$ dependence} \label{sec:scale_sep}

In \citetalias{2024MNRAS.530.1866S} we found that the cutoff energy for accelerated particles is proportional to $\omega_{c,0}/\Omega_0$. In order to understand the possible applications of this result to real systems, the dependence of the different acceleration stages identified in this work on the scale separation parameter $\omega_{c,0}/\Omega_0$ needs to be determined. 

\begin{figure*}%[ht!]
    \centering
    \includegraphics[width=.85\textwidth]{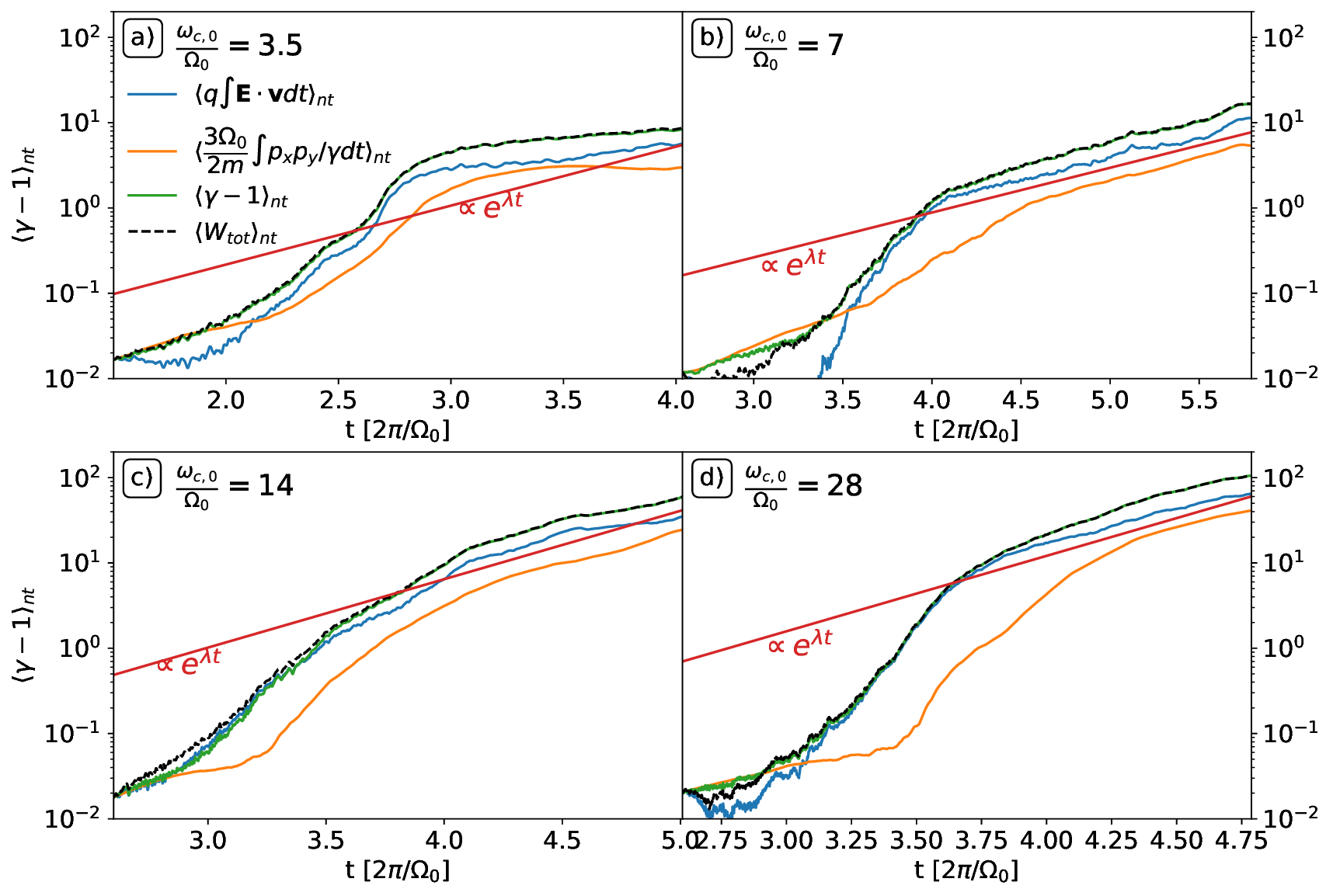}
    \caption{Same as panel $a$ of Figure \ref{fig:meanE_simt}, for simulations with scale separations $\omega_{c,0}/\Omega_0=3.5, 7, 14,$ and $28$ in panels $a, b, c,$ and $d$, respectively.}
    \label{fig:acc_scale}
\end{figure*}

For simplicity, let us begin with the final stage. Panels $a$-$d$ of Figure \ref{fig:acc_scale} show the average energy evolution of the tracked particles that reach energies larger than the cut-off energy and are within the disk region at the end of the simulation, which we define as the moment when ${k_B\overline{T} \approx 0.3\, mc^2}$, 
for simulations with scale separations $\omega_{c,0}/\Omega_0=3.5, 7, 14,$ and $28$ (see in Table \ref{tab:cut_off} the values of the cutoff Lorentz factors $\gamma_c$ for each simulation). The average particle energy ${\langle \gamma - 1 \rangle_{nt}}$ is shown in green, the shear acceleration contribution is in orange, the electric field's work contribution is in blue, and the sum of them is the black-dashed line. The solid red line shows the prediction from second-order Fermi acceleration, computed for each simulation using its estimated turbulent velocity, which gives an acceleration rate $\lambda = 0.25,\,0.19,\,0.29,$ and $0.32\, \Omega_0$ for the runs with scale separations $\omega_{c,0}/\Omega_0=3.5, 7, 14,$ and $28$, respectively. 
While most of the energy gain comes from work done by the electric field, shear acceleration contributes $30\%-40\%$, regardless of the scale separation. 
At large enough scales separation $\omega_{c,0}/\Omega_0\gtrsim 7$, the electric field contribution is well described by the second-order Fermi prediction shown by the solid red line. Therefore, as long as ${\omega_{c,0}/\Omega_0 \geq 7}$, at late times $t\gtrsim 4\,[2\pi/\Omega_0]$ the tracked particles are mainly experiencing second-order Fermi acceleration with significant contribution ($\sim 30\%$) from shear acceleration.

Regarding applications to real accretion disks, this  second-order Fermi acceleration should be limited by the accretion time, $t_a$, onto the black hole. Indeed, it can be shown that $\Omega_0t_a\approx 2\alpha^{-1}$ \citep{1997ApJ...490..605M}, where $\alpha$ is the viscosity parameter of the disk \citep{1973A&A....24..337S}. In our simulations, we have demonstrated that in the disk $\alpha \approx 0.5$ \citepalias{2024MNRAS.530.1866S}, fairly independently of the scale separation ratio $\omega_{c,0}/\Omega_0$, implying that second-order Fermi acceleration has a time of the order of a few $\Omega_0^{-1}$ to act. This means that second-order Fermi acceleration (and shear) can only increase the non-thermal particles' energy by a fixed factor of the order of $\sim e^{\lambda t_a} \sim \exp(1.2)$ (with $\lambda \approx 0.3\Omega_0$). This time limitation also applies to our simulations, where this second-order Fermi-dominated stage only last for about one orbital period ($2\pi/\Omega_0$). This means that the dependence of $\gamma_c \propto \omega_{c,0}/\Omega_0$ found in \citetalias{2024MNRAS.530.1866S} must be explained by the previous two stages.%This implies that the energy gain during reconnection-driven acceleration is what controls the maximum particle energy, which is only enhanced by a factor of a few by second-order Fermi plus shear acceleration.

\begin{figure*}%[ht!]
    \centering
    \includegraphics[width=.85\textwidth]{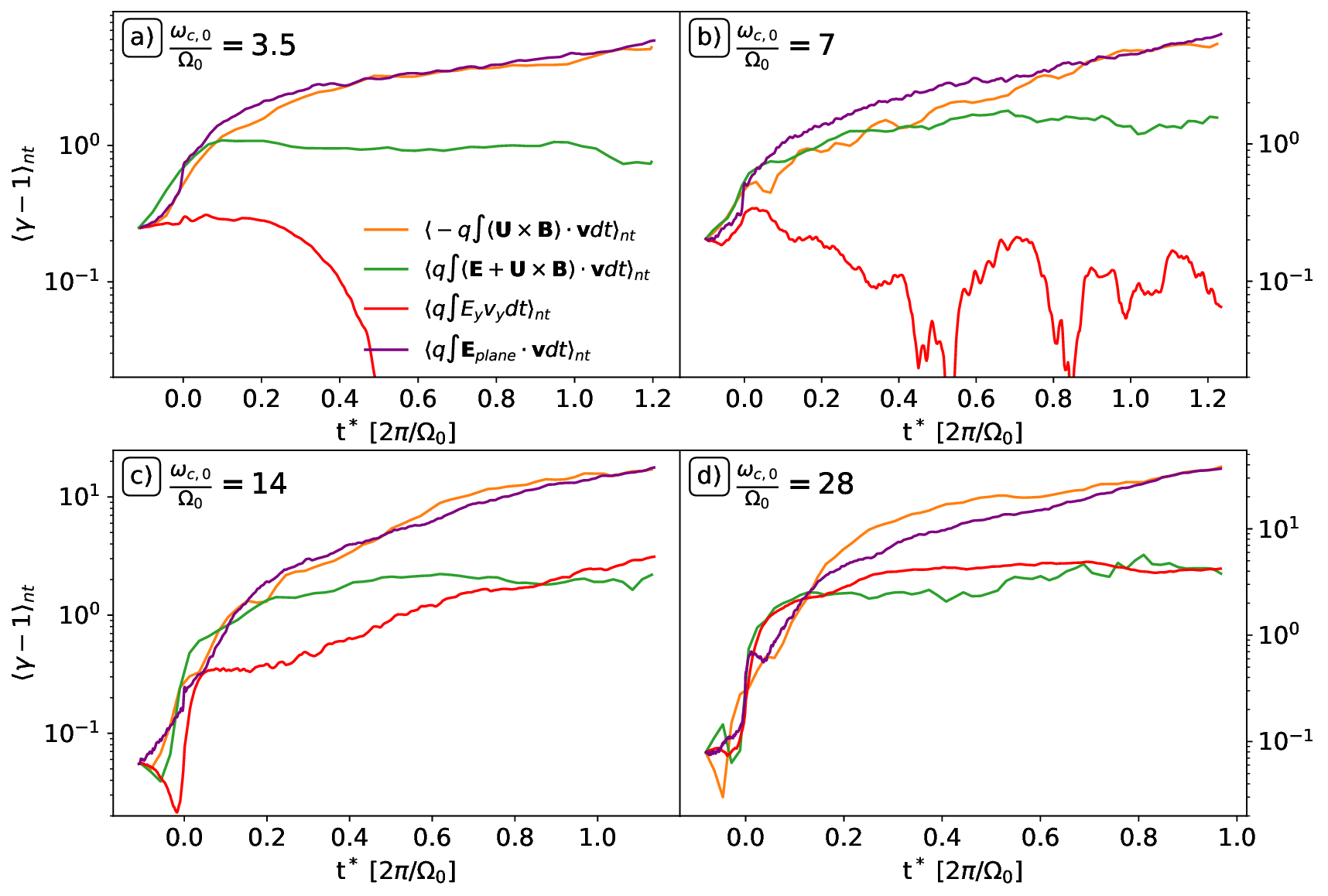}
    \caption{Same as panel $b$ of Figure \ref{fig:injection_28}, for simulations with scale separations $\omega_{c,0}/\Omega_0=3.5, 7, 14,$ and $28$ in panels $a, b, c,$ and $d$, respectively.}
    \label{fig:inj_scale}
\end{figure*}

In order to determine the role played by $\omega_{c,0}/\Omega_0$ on the initial injection and the subsequent first-order Fermi-like acceleration, Figure \ref{fig:inj_scale} shows the evolution of the electric field's work divided into ideal and non-ideal components, for the same simulations analyzed in Fig. \ref{fig:acc_scale}. This figure also includes the contributions along the azimuthal axis and in the poloidal plane, as a function of the post-injection time $t^*$.\footnote{The time range is so that at least $85\%$ of the selected particles are included in the analysis at each time.} For all scale separations, at the injection time $t^*=0$ the non-ideal part of the electric field presents a rapid rise in its contribution to the energy of the particles, which dominates the initial energization of the particles.

This is followed by a more slowly growing contribution from the ideal electric field, which eventually becomes dominant. In the case of $\omega_{c,0}/\Omega_0 = 28$ analyzed in Section~\ref{sec:inj}, we identified this ideal field-dominated stage with particles following Speiser-like trajectories along the current sheets formed by the large-scale, dynamo-driven magnetic field. In particular, we associated the ideal energy gain with vertically converging plasma motions in the region between plasmoids that move away from each other (after being formed by the splitting of a larger, plasmoid, as shown in Fig.\ref{fig:trajectory_zoom}), giving rise to a first-order Fermi-like acceleration. In this first-order Fermi scenario, the time derivative of the Lorentz factor of a non-thermal particle should satisfy \citep{2011ApJ...737L..40U,2012ApJ...746..148C}
\begin{equation}
    \dot{\gamma} \sim \eta \omega_c,
\end{equation}
where $\omega_c$ is its non-relativistic cyclotron frequency and $\eta$ quantifies the motional electric field as 

\begin{equation}
    \eta = \frac{|\mathbf{U}\times\mathbf{B}|}{B c}.
\end{equation} 

\begin{figure}%[ht!]
    \centering
    \includegraphics[width=\columnwidth]{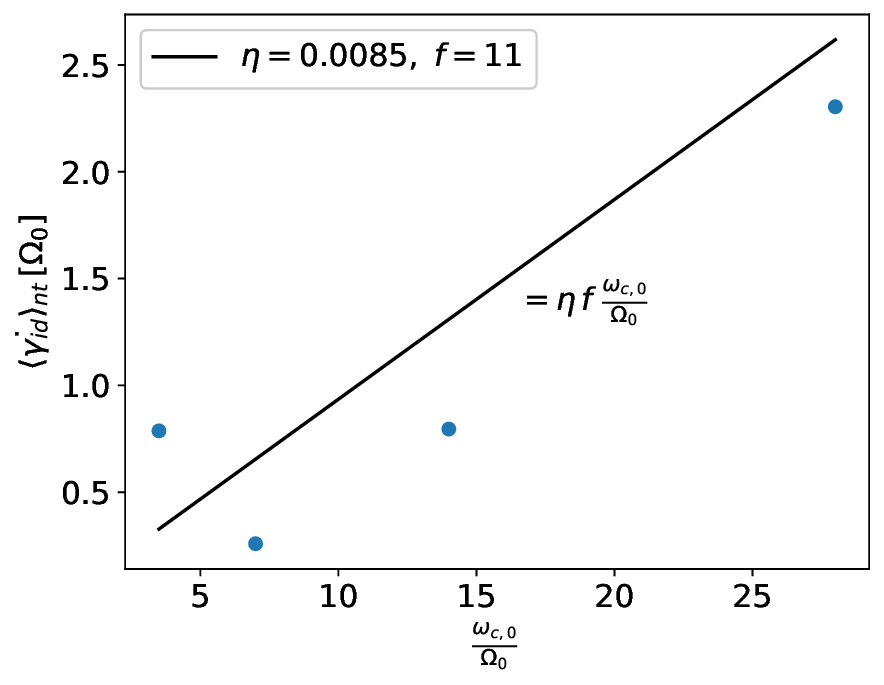}
    \caption{Power contribution of the ideal electric field to the energy of the particles during the transition from the non-ideal dominated injection to the ideal dominated phase. The linear growth is deduced from a magnetic field amplification factor $f=11$ and a projected ideal electric field efficiency ${\eta\equiv \frac{1}{\langle v \rangle_{nt}}\langle \frac{-q(\mathbf{U}\times \mathbf{B})\cdot\mathbf{v}}{B}\rangle_{nt}=0.0085}$.}
    \label{fig:Ideal_power}
\end{figure}

\begin{figure*}%[ht!]
    \centering
    \includegraphics[width=.85\textwidth]{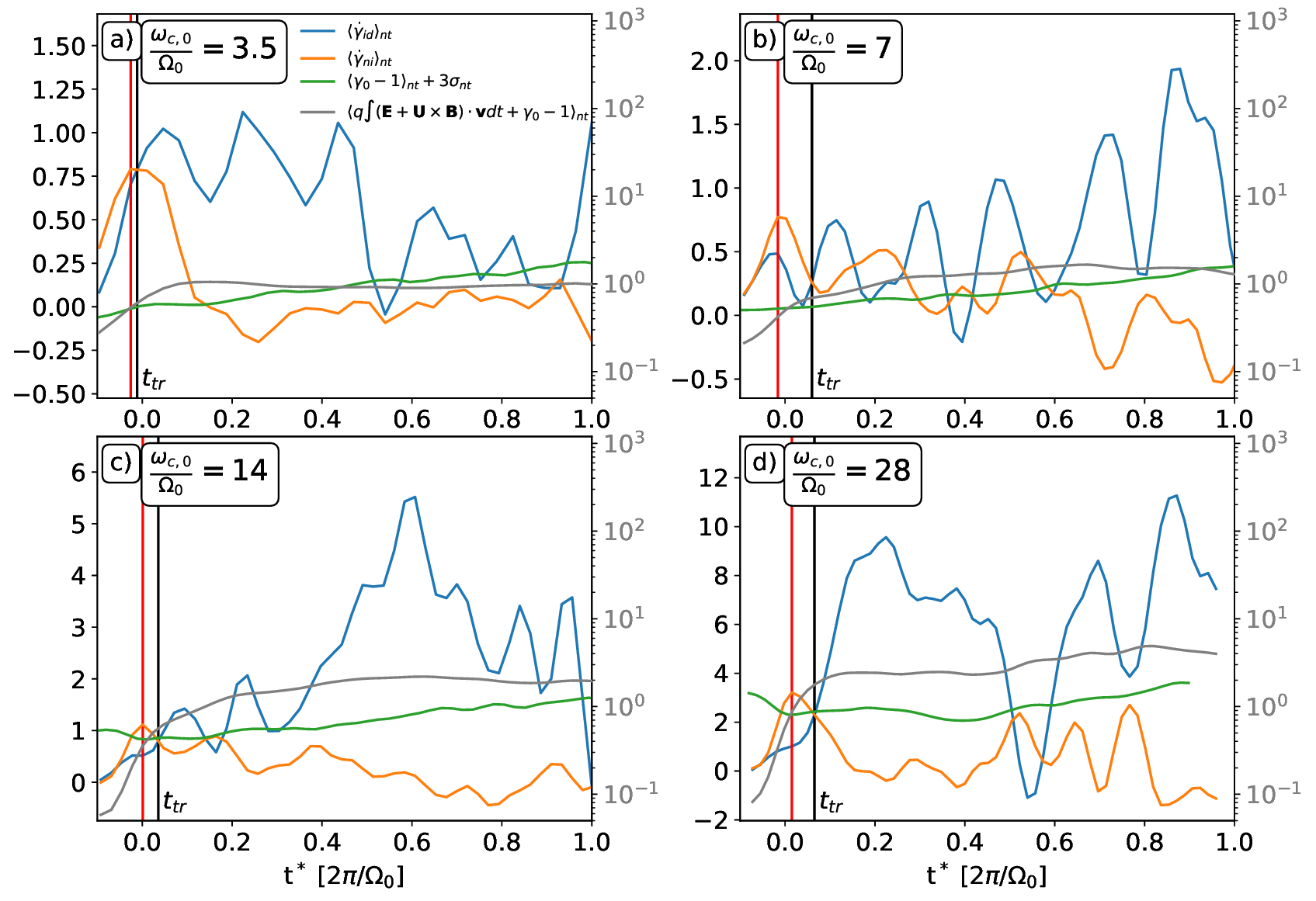}
    \caption{Panels $a$-$d$ show, for different scale separations $\omega_{c,0}/\Omega_0$ the power done by the ideal ($\dot{\gamma}_{id}$, blue) and the non-ideal ($\dot{\gamma}_{ni}$, orange) electric field on the left axes. The right axes show the non-ideal electric field work (gray); and the sum of the initial particle's energy $\langle \gamma_0-1\rangle_{nt}$ and $3\sigma_{nt}$ (green), where $\sigma_{nt}\equiv\frac{\langle B^2\rangle_{nt}}{\langle4\pi nmc^2\rangle_{nt}}$ is computed at the position of each particle and $\gamma_0-1$ is the particle's energy at time $t^*\approx -0.1 (2\pi/\Omega_0)$. The vertical red and black lines show the times when the non-ideal power is maximum and the transition from non-ideal to ideal dominated acceleration.}
    \label{fig:non-ideal_cont}
\end{figure*}

If this process applies to all our tested values of $\omega_{c,0}/\Omega_0$, the time derivative of $\gamma$ normalized by $\Omega_0$ should be proportional to $\omega_{c,0}/\Omega_0$, which requires that both $\eta$ and the average amplification factor $f=\omega_c/\omega_{c,0}$ are independent of $\omega_{c,0}/\Omega_0$. This is indeed the case for simulations with large enough scale separations, in which $\omega_{c,0}/\Omega_0\gtrsim 7$. Figure \ref{fig:Ideal_power} shows the acceleration power from the ideal electric field work $\langle\dot{\gamma_{id}}\rangle_{nt}$, i.e., the average over all the accelerated particles for $\dot{\gamma}=-\frac{q}{mc^2}(\mathbf{U}\times\mathbf{B})\cdot\mathbf{v}$, at the transition from the non-ideal reconnection-driven injection to the ideal first-order Fermi-like stage, which is determined by finding the time at which the power from the ideal and non-ideal fields become equal.\footnote{These times are marked by vertical black lines in Fig. \ref{fig:non-ideal_cont}, which shows in orange and blue the average time derivative of the nonthermal particles' Lorentz factors due to non-ideal and ideal electric field, respectively. These time derivatives are calculated by previously smoothing the work shown in Figure \ref{fig:inj_scale} using a Gaussian filter with window ${0.026\, [2\pi/\Omega_0]}$.} We see that for $\omega_{c,0}/\Omega_0\geq 7$, ${\langle\dot{\gamma}\rangle_{nt}\,[\Omega_0]}$ is fairly well described by a linear function of ${\omega_{c,0}/\Omega_0}$, approaching 
\begin{equation}
\frac{\langle \dot{\gamma}\rangle_{nt}}{\Omega_0} = \eta f \omega_{c,0}/\Omega_0,
\label{eq:eficiencia}
\end{equation}
which is shown by the solid black line in Fig~\ref{fig:Ideal_power}, using $\eta=0.0085$ and $f=11$, where $f$ is the amplification factor of the magnetic field in the disk (i.e., $f\equiv \langle B^2\rangle_v^{1/2}/B_0$). These values for $f$ and $\eta$ represent fairly well the $f$ and $\eta$ parameters in the simulations with $\omega_{c,0}/\Omega_0 \geq 7$ at the time when ideal and non-ideal powers become equal (with an error of $\sim25\%$).
This result indicates that for $\omega_{c,0}/\Omega_0\geq7$, this acceleration is fast, with the time derivative of the accelerated particles being $\eta \omega_c$, where $\omega_c$ is the instantaneous non-relativistic cyclotron frequency of particles.

Regarding the maximum particle energy, this implies that, as $\omega_{c,0}/\Omega_0$ increases, in the time taken by a particle to cross the space between plasmoids \citepalias[which grows proportional to $\omega_{c,0}/\Omega_0$ as we saw in ][]{2024MNRAS.530.1866S} the accelerated particles will have time to reach proportionally larger energies. More precisely, the maximum energy should be set by balancing the acceleration time ${t_{acc}=\gamma/\dot{\gamma} \sim \gamma/\eta\omega_c}$ with the crossing time between plasmoids ${t_{cr} \sim L/c}$. Estimating $L$ as the dominant length-scale within the MRI turbulence, which is ${L \approx (2\pi)^2v_{A,0}/\Omega_0}$ fairly independently of $\omega_{c,0}/\Omega_0$ \citepalias{2024MNRAS.530.1866S} and using that in our runs $v_{A,0}/c=10^{-2}$, we obtain that the maximum Lorentz factor attainable during the first-order Fermi-like acceleration stage is
\begin{equation}
\gamma_{Fermi I} \sim 0.04 \frac{\omega_{c,0}}{\Omega_0},
\label{eq:scaling}
\end{equation}
where we have used the fact that in our runs with $\omega_{c,0}/\Omega_0 >= 7$, $\eta \approx 0.01$ and the magnetic field in the disk has been amplified by a factor $f\approx 10$. 

This result is important because, by combining Eq. \ref{eq:scaling} with the fact that the second-order Fermi-dominated stage only increases the energy of particles by a factor of {\it a few} that is independent of $\omega_{c,0}/\Omega_0$, we obtain that $\gamma_c$ should scale as $\gamma_c \sim \omega_{c,0}/\Omega_0$, consistent with \citetalias{2024MNRAS.530.1866S}. Regarding the initial (non-ideal) reconnection-driven injection, we check its dependence on $\omega_{c,0}/\Omega_0$. Using Fig. \ref{fig:non-ideal_cont}, where we compare the non-ideal field work with $3\sigma_{nt}$, where $\sigma_{nt}$ is defined as ${\langle B^2 \rangle_{nt}/\langle 4\pi n mc^2\rangle_{nt}}$ (the averages $\langle \, \rangle_{nt}$ are taken at the non-thermal particles' positions). For all the scale separations used, we see that from $t^*=-0.1$ to the moment where the power of the non-ideal field is maximum the work done from the non-ideal electric field satisfy $ {\langle \frac{q}{mc^2}\int(\mathbf{E}+\mathbf{U}\times\mathbf{B})\cdot\mathbf{v}dt^* \rangle_{nt} \approx 3\sigma_{nt}}$, which is itself fairly independent of $\omega_{c,0}/\Omega_0$, implying that scale-separation does not play a role in the energy acquired by particles in the initial injection process.

\begin{figure*}%[ht!]
    \centering
    \includegraphics[width=.85\textwidth]{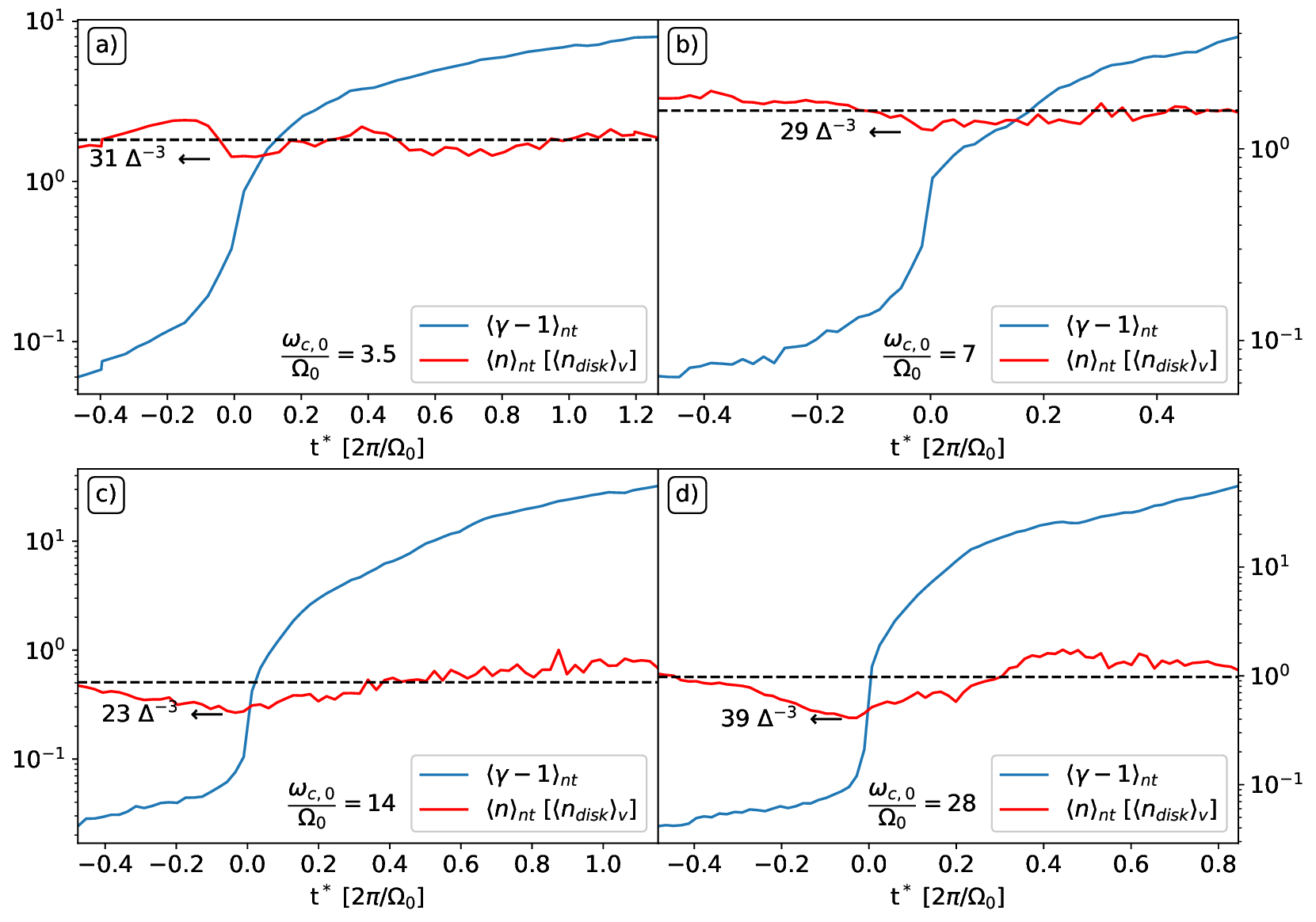}
    \caption{Same as Figure \ref{fig:density}, for simulations with scale separations $\omega_{c,0}/\Omega_0=3.5, 7, 14,$ and $28$ in panels $a, b, c,$ and $d$, respectively. The horizontal arrow from the minimum shows the average minimum density as the average number of particles in the simulation cell.}
    \label{fig:dens_scale}
\end{figure*}

Regarding the dependence of the injection dynamics on scale-separation, as shown for $\omega_{c,0}/\Omega_0=28$, during the initial injection process particles escape from high density regions to lower density regions, as a consequence of a reconnection event. Figure \ref{fig:dens_scale} shows the evolution of the particle-averaged background plasma density, $\langle n\rangle_{nt}$ in units of the volume-average disk density, $\langle n_{disk} \rangle_v$ using the post-injection time $t^*$, for simulations with different scale separations. The background density shows a local minimum at the moment of the injection $t^*=0$, with the contrast between the minimum and the average density (horizontal black dashed lines) increasing with the scale separation. This suggests that, as more realistic scale separations are used, the escaping of the non-thermal particles from splitting plasmoids at injection becomes more robust. 

In summary, while the injection yields energies that are fairly independent of $\omega_{c,0}/\Omega_0$, the first order Fermi-like process clearly shows an energy increase that is proportional to the scale separation, while the final second-order Fermi-dominated process only provides a fairly scale separation-independent, order unity factor to the maximum energy. Putting these three stages together, this results in an acceleration that produces maximum energies that are proportional to $\omega_{c,0}/\Omega_0$, consistently with the $\gamma_c \sim \omega_{c,0}/\Omega_0$ result found by \citepalias{2024MNRAS.530.1866S}.

\section{Conclusions} \label{sec:conc}

Earlier 2D PIC simulations of the stratified MRI showed that, during the turbulent evolution of the MRI, the particle energy distributions develop a non-thermal, high-energy tail that is well-fitted by a power law with an exponential cutoff (as shown in Figure \ref{fig:spect28}), with the cutoff Lorentz factor, $\gamma_c$, satisfying $\gamma_c \sim \omega_{c,0}/\Omega_0$ \citepalias{2024MNRAS.530.1866S}. In this work, we disentangle the acceleration mechanisms involved by tracking the evolution of a subpopulation of accelerated particles that reach at least the cutoff Lorentz factor $\gamma_c$ by the end of the simulation, defined as the time when the disk temperature reaches $0.3\,mc^2$ (the values of $\gamma_c$ for simulations with different $\omega_{c,0}/\Omega_0$ are shown in Table \ref{tab:cut_off}). Since the values of $\omega_{c,0}/\Omega_0$ used in our runs are much smaller than those in realistic astrophysical settings, our analysis pays particular attention to the efficiency of the identified acceleration mechanisms and their dependence on $\omega_{c,0}/\Omega_0$.

Overall, we find that acceleration occurs mainly in regions where the nearly azimuthal magnetic field produced by a large-scale MRI dynamo changes polarity, as shown in Figure \ref{fig:dynamo_inj}. The acceleration is dominated by work done by the electric field produced by the MRI turbulence, with a subdominant contribution of about $30\%$ from global shear acceleration, regardless of the scale separation, as shown in panels ($a$)-($d$) of Figure \ref{fig:acc_scale}.

We find that the acceleration can be divided into three stages: an initial stage in which particles are injected by magnetic reconnection, a second stage in which acceleration is driven by a first-order Fermi-like process, and a final stage dominated by second-order Fermi acceleration, with a subdominant contribution from shear acceleration. Although most of the acceleration is produced by ideal electric fields, the initial reconnection-driven injection is dominated by non-ideal electric fields. This injection is characterized by rapid energization that raises the particle energy well above the thermal energy, reaching $\langle\gamma\rangle_{nt} \approx 3{\sigma}_{nt}$ (${\sigma}_{nt} \equiv \langle B^2\rangle_{nt}/\langle 4\pi nmc^2\rangle_{nt}$, where $\langle\,\rangle_{nt}$ denotes an average over the positions of the non-thermal particles), independently of $\omega_{c,0}/\Omega_0$, as shown in Figure \ref{fig:non-ideal_cont}. This behavior is consistent with expectations from reconnection \citep[e.g., ][]{2021ApJ...922..261Z,2023ApJ...956L..36Z}. Indeed, tracking individual non-thermal particles in the case $\omega_{c,0}/\Omega_0=28$ shows that this injection process results from reconnection of the poloidal field components (the only components allowed to reconnect in 2D) during the splitting of plasmoids produced by the MRI turbulence. This splitting occurs because the plasmoids are not static but move while merging and stretching, occasionally breaking apart. Rather than becoming trapped by the smaller, newly formed plasmoids, the injected particles escape into the low-density region between them. This escape into the lower-density regions outside the plasmoids becomes clearer as the scale separation increases, as shown in Figure \ref{fig:dens_scale}.

Our tracking of individual particles shows that, once outside the plasmoids, the non-thermal particles enter a second acceleration stage in which they are further energized by the ideal convective electric field produced by vertically converging plasma motions, which gives rise to a first-order Fermi-like acceleration. This acceleration happens while the particles undergo Speiser-like trajectories along the current sheet formed by the large-scale, dynamo-driven magnetic fields. These trajectories occur between the newly formed plasmoids, near the X-point of the reconnection event that produced them, as the plasmoids move apart. This acceleration is rapid, with $\dot{\gamma} \approx \eta\omega_c$, where $\eta$ quantifies the motional electric field relative to the local magnetic field. In Figure \ref{fig:Ideal_power}, we show that this rapid-acceleration scenario characterizes the entire population of non-thermal particles immediately after injection and holds for runs with $\omega_{c,0}/\Omega_0 \geq 7$, for which $\eta \approx 0.01$. This property of the identified first-order Fermi-like mechanism implies that particles reach energies proportional to $\omega_{c,0}/\Omega_0$.

In addition to these two initial acceleration stages, there is a final stage in which energization is dominated by second-order Fermi acceleration and, to a lesser extent, by shear acceleration. During this stage, the energy of the non-thermal particles increases exponentially at a rate $\lambda \approx 0.3\,\Omega_0$. Since this final stage lasts for about one orbit $\sim 2\pi/\Omega_0$ in our simulations, it should increase the maximum energy by only a factor of {\it a few}, regardless of $\omega_{c,0}/\Omega_0$, thereby preserving the proportionality between the particle energy gain and $\omega_{c,0}/\Omega_0$, as observed for the cut-off energy attained in the simulations of \citetalias{2024MNRAS.530.1866S}. In realistic settings, the available acceleration time should remain of order $\Omega_0^{-1}$ as it is set by the accretion time, which is also of order $\Omega_0^{-1}$ for a viscosity parameter $\alpha \approx 0.5$, as found in \citetalias{2024MNRAS.530.1866S}. Thus, the identified sequence of acceleration stages is consistent with and explains the relation $\gamma_c \sim \omega_{c,0}/\Omega_0$ found in \citetalias{2024MNRAS.530.1866S}.

The consistency of the identified acceleration process with a maximum particle energy proportional to $\omega_{c,0}/\Omega_0$ makes the MRI a promising candidate for accelerating particles to ultrarelativistic energies in weakly collisional accretion disks around black holes. However, caution is warranted since our results apply to 2D simulations with $m_i/m_e=1$. Further research is therefore needed to extend them to 3D, where the azimuthal component of the magnetic field can reconnect, and to $m_i/m_e>1$. Nevertheless, this work constitutes a valuable step toward identifying the acceleration mechanisms operating in the collisionless MRI and determining their dependence on the scale separation $\omega_{c,0}/\Omega_0$, a key aspect to understand the possible application of these results to realistic accretion disks around black holes.

%\section{acknowledgments}
We thank Elisabete M. de Gouveia Dal Pino, Anatoly Spitkovsky, Fabio Bacchini, and Martin Lemoine for useful comments and discussion. 
AS and JC acknowledge ANID through Millennium Science Initiative Program NCN2023\_002.
AS and MR acknowledge the Center for Astrophysics and Associated Technologies (CATA; ANID Basal grant FB210003).
MR gratefully acknowledges support from the ANID-FONDECYT grant 1260829.
JC acknowledges the financial support from ANID-FONDECYT Regular 1251444.
The computational resources and services used in this work were provided by the supercomputing infrastructure of the NLHPC (CCSS210001), at the Center for Mathematical Modeling of University of Chile.

%% For this sample we use BibTeX plus aasjournalv7.bst to generate the

\bibliography{refs}{}

@ARTICLE{2012ApJ...746..148C,
       author = {{Cerutti}, Beno{\^\i}t and {Uzdensky}, Dmitri A. and {Begelman}, Mitchell C.},
        title = "{Extreme Particle Acceleration in Magnetic Reconnection Layers: Application to the Gamma-Ray Flares in the Crab Nebula}",
      journal = {\apj},
         year = 2012,
        month = feb,
       volume = {746},
       number = {2},
          eid = {148},
        pages = {148},
          doi = {10.1088/0004-637X/746/2/148},
archivePrefix = {arXiv},
       eprint = {1110.0557},
 primaryClass = {astro-ph.HE},
       adsurl = {https://ui.adsabs.harvard.edu/abs/2012ApJ...746..148C}
}

@ARTICLE{2026arXiv260401222D,
       author = {{Das}, Abhishek and {Murase}, Kohta and {Zhang}, B. Theodore},
        title = "{Multimessenger Constraints on Production Sites of High-Energy Neutrinos from NGC 1068}",
      journal = {arXiv e-prints},
         year = 2026,
        month = apr,
          eid = {arXiv:2604.01222},
        pages = {arXiv:2604.01222},
          doi = {10.48550/arXiv.2604.01222},
archivePrefix = {arXiv},
       eprint = {2604.01222},
 primaryClass = {astro-ph.HE},
       adsurl = {https://ui.adsabs.harvard.edu/abs/2026arXiv260401222D}
}

@ARTICLE{2026arXiv260215644E,
       author = {{Eichmann}, Bj{\"o}rn and {Salvatore}, Silvia and {del Palacio}, Santiago and {Sommani}, Giacomo and {Mele}, Crystal and {Veres}, Patrik M. and {Becker Tjus}, Julia},
        title = "{Bayesian parameter study of the Seyfert-starburst composite galaxies NGC 1068 and NGC 7469}",
      journal = {arXiv e-prints},
         year = 2026,
        month = feb,
          eid = {arXiv:2602.15644},
        pages = {arXiv:2602.15644},
          doi = {10.48550/arXiv.2602.15644},
archivePrefix = {arXiv},
       eprint = {2602.15644},
 primaryClass = {astro-ph.HE},
       adsurl = {https://ui.adsabs.harvard.edu/abs/2026arXiv260215644E}
}

@ARTICLE{2023ApJ...956....8F,
       author = {{Fang}, Ke and {Lopez Rodriguez}, Enrique and {Halzen}, Francis and {Gallagher}, John S.},
        title = "{High-energy Neutrinos from the Inner Circumnuclear Region of NGC 1068}",
      journal = {\apj},
         year = 2023,
        month = oct,
       volume = {956},
       number = {1},
          eid = {8},
        pages = {8},
          doi = {10.3847/1538-4357/acee70},
archivePrefix = {arXiv},
       eprint = {2307.07121},
 primaryClass = {astro-ph.HE},
       adsurl = {https://ui.adsabs.harvard.edu/abs/2023ApJ...956....8F}
}

@ARTICLE{2025ApJ...990..170X,
       author = {{Xue}, Rui and {He}, Jia-Chun and {Xiong}, Dingrong and {Wang}, Ze-Rui},
        title = "{On the Origin of a Possible Hard Very High Energy Spectrum from M87 Discovered by LHAASO}",
      journal = {\apj},
         year = 2025,
        month = sep,
       volume = {990},
       number = {2},
          eid = {170},
        pages = {170},
          doi = {10.3847/1538-4357/adf72c},
archivePrefix = {arXiv},
       eprint = {2508.01986},
 primaryClass = {astro-ph.HE},
       adsurl = {https://ui.adsabs.harvard.edu/abs/2025ApJ...990..170X}
}

@ARTICLE{2026ApJ...997..151M,
       author = {{Mondal}, Nibedita and {Mondal}, Sandeep Kumar and {Gupta}, Nayantara},
        title = "{Explaining the Origin of TeV Gamma Rays from M87 during High and Low States}",
      journal = {\apj},
         year = 2026,
        month = feb,
       volume = {997},
       number = {2},
          eid = {151},
        pages = {151},
          doi = {10.3847/1538-4357/ae1f0b},
archivePrefix = {arXiv},
       eprint = {2509.21314},
 primaryClass = {astro-ph.HE},
       adsurl = {https://ui.adsabs.harvard.edu/abs/2026ApJ...997..151M}
}

@article{1965JGR....70.4219S,
  author  = {{Speiser}, T.~W.},
  title   = {{Particle Trajectories in Model Current Sheets, 1, Analytical Solutions}},
  journal = {\jgr},
  year    = 1965,
  month   = sep,
  volume  = {70},
  number  = {17},
  pages   = {4219-4226},
  doi     = {10.1029/JZ070i017p04219},
  adsurl  = {https://ui.adsabs.harvard.edu/abs/1965JGR....70.4219S}
}

@article{1973A&A....24..337S,
  author  = {{Shakura}, N.~I. and {Sunyaev}, R.~A.},
  title   = {{Black holes in binary systems. Observational appearance.}},
  journal = {\aap},
  year    = 1973,
  month   = jan,
  volume  = {24},
  pages   = {337-355},
  adsurl  = {https://ui.adsabs.harvard.edu/abs/1973A&A....24..337S}
}

@article{1981ARA&A..19..137P,
  author   = {{Pringle}, J.~E.},
  title    = {{Accretion discs in astrophysics}},
  journal  = {\araa},
  year     = 1981,
  month    = jan,
  volume   = {19},
  pages    = {137-162},
  doi      = {10.1146/annurev.aa.19.090181.001033},
  adsurl   = {https://ui.adsabs.harvard.edu/abs/1981ARA&A..19..137P}
}

@article{1981SvAL....7..352B,
  author  = {{Berezhko}, E.~G. and {Krymskii}, G.~F.},
  title   = {{a Kinetic Analysis of the Charged Particle Acceleration Process in Collisionless Plasma Shear Flows}},
  journal = {Soviet Astronomy Letters},
  year    = 1981,
  month   = jun,
  volume  = {7},
  pages   = {352},
  adsurl  = {https://ui.adsabs.harvard.edu/abs/1981SvAL....7..352B}
}

@article{1988ApJ...331L..91E,
  author   = {{Earl}, J.~A. and {Jokipii}, J.~R. and {Morfill}, G.},
  title    = {{Cosmic-Ray Viscosity}},
  journal  = {\apjl},
  year     = 1988,
  month    = aug,
  volume   = {331},
  pages    = {L91},
  doi      = {10.1086/185242},
  adsurl   = {https://ui.adsabs.harvard.edu/abs/1988ApJ...331L..91E}
}

@article{1991ApJ...376..214B,
  author   = {{Balbus}, Steven A. and {Hawley}, John F.},
  title    = {{A Powerful Local Shear Instability in Weakly Magnetized Disks. I. Linear Analysis}},
  journal  = {\apj},
  year     = 1991,
  month    = jul,
  volume   = {376},
  pages    = {214},
  doi      = {10.1086/170270},
  adsurl   = {https://ui.adsabs.harvard.edu/abs/1991ApJ...376..214B}
}

@article{1995ApJ...440..742H,
  author   = {{Hawley}, John F. and {Gammie}, Charles F. and {Balbus}, Steven A.},
  title    = {{Local Three-dimensional Magnetohydrodynamic Simulations of Accretion Disks}},
  journal  = {\apj},
  year     = 1995,
  month    = feb,
  volume   = {440},
  pages    = {742},
  doi      = {10.1086/175311},
  adsurl   = {https://ui.adsabs.harvard.edu/abs/1995ApJ...440..742H}
}

@article{1997ApJ...489..865E,
  author        = {{Esin}, Ann A. and {McClintock}, Jeffrey E. and {Narayan}, Ramesh},
  title         = {{Advection-Dominated Accretion and the Spectral States of Black Hole X-Ray Binaries: Application to Nova Muscae 1991}},
  journal       = {\apj},
  year          = 1997,
  month         = nov,
  volume        = {489},
  number        = {2},
  pages         = {865-889},
  doi           = {10.1086/304829},
  archiveprefix = {arXiv},
  eprint        = {astro-ph/9705237},
  primaryclass  = {astro-ph},
  adsurl        = {https://ui.adsabs.harvard.edu/abs/1997ApJ...489..865E}
}

@article{1997ApJ...490..605M,
  author        = {{Mahadevan}, Rohan and {Quataert}, Eliot},
  title         = {{Are Particles in Advection-dominated Accretion Flows Thermal?}},
  journal       = {\apj},
  year          = 1997,
  month         = dec,
  volume        = {490},
  number        = {2},
  pages         = {605-618},
  doi           = {10.1086/304908},
  archiveprefix = {arXiv},
  eprint        = {astro-ph/9705067},
  primaryclass  = {astro-ph},
  adsurl        = {https://ui.adsabs.harvard.edu/abs/1997ApJ...490..605M}
}

@article{1998RvMP...70....1B,
  author   = {{Balbus}, Steven A. and {Hawley}, John F.},
  title    = {{Instability, turbulence, and enhanced transport in accretion disks}},
  journal  = {Reviews of Modern Physics},
  year     = 1998,
  month    = jan,
  volume   = {70},
  number   = {1},
  pages    = {1-53},
  doi      = {10.1103/RevModPhys.70.1},
  adsurl   = {https://ui.adsabs.harvard.edu/abs/1998RvMP...70....1B}
}

@article{2003ApJ...598..301Y,
  author        = {{Yuan}, Feng and {Quataert}, Eliot and {Narayan}, Ramesh},
  title         = {{Nonthermal Electrons in Radiatively Inefficient Accretion Flow Models of Sagittarius A*}},
  journal       = {\apj},
  year          = 2003,
  month         = nov,
  volume        = {598},
  number        = {1},
  pages         = {301-312},
  doi           = {10.1086/378716},
  archiveprefix = {arXiv},
  eprint        = {astro-ph/0304125},
  primaryclass  = {astro-ph},
  adsurl        = {https://ui.adsabs.harvard.edu/abs/2003ApJ...598..301Y}
}

@article{2005A&A...441..845D,
  author   = {{de Gouveia dal Pino}, E.~M. and {Lazarian}, A.},
  title    = {{Production of the large scale superluminal ejections of the microquasar GRS 1915+105 by violent magnetic reconnection}},
  journal  = {\aap},
  year     = 2005,
  month    = oct,
  volume   = {441},
  number   = {3},
  pages    = {845-853},
  doi      = {10.1051/0004-6361:20042590},
  adsurl   = {https://ui.adsabs.harvard.edu/abs/2005A&A...441..845D}
}

@inproceedings{2005AIPC..801..345S,
  author        = {{Spitkovsky}, Anatoly},
  title         = {{Simulations of relativistic collisionless shocks: shock structure and particle acceleration}},
  booktitle     = {Astrophysical Sources of High Energy Particles and Radiation},
  year          = 2005,
  editor        = {{Bulik}, Tomasz and {Rudak}, Bronislaw and {Madejski}, Grzegorz},
  series        = {American Institute of Physics Conference Series},
  volume        = {801},
  month         = nov,
  publisher     = {AIP},
  pages         = {345-350},
  doi           = {10.1063/1.2141897},
  archiveprefix = {arXiv},
  eprint        = {astro-ph/0603211},
  primaryclass  = {astro-ph},
  adsurl        = {https://ui.adsabs.harvard.edu/abs/2005AIPC..801..345S}
}

@article{2006ApJ...636..777A,
  author        = {{Aharonian}, F. and {Akhperjanian}, A.~G. and {Bazer-Bachi}, A.~R. and {Beilicke}, M. and {Benbow}, W. and {Berge}, D. and {Bernl{\"o}hr}, K. and {Boisson}, C. and {Bolz}, O. and {Borrel}, V. and {Braun}, I. and {Breitling}, F. and {Brown}, A.~M. and {Chadwick}, P.~M. and {Chounet}, L. -M. and {Cornils}, R. and {Costamante}, L. and {Degrange}, B. and {Dickinson}, H.~J. and {Djannati-Ata{\"\i}}, A. and {Drury}, L. O'C. and {Dubus}, G. and {Emmanoulopoulos}, D. and {Espigat}, P. and {Feinstein}, F. and {Fontaine}, G. and {Fuchs}, Y. and {Funk}, S. and {Gallant}, Y.~A. and {Giebels}, B. and {Gillessen}, S. and {Glicenstein}, J.~F. and {Goret}, P. and {Hadjichristidis}, C. and {Hauser}, M. and {Heinzelmann}, G. and {Henri}, G. and {Hermann}, G. and {Hinton}, J.~A. and {Hofmann}, W. and {Holleran}, M. and {Horns}, D. and {Jacholkowska}, A. and {de Jager}, O.~C. and {Kh{\'e}lifi}, B. and {Komin}, Nu. and {Konopelko}, A. and {Latham}, I.~J. and {Le Gallou}, R. and {Lemi{\`e}re}, A. and {Lemoine-Goumard}, M. and {Leroy}, N. and {Lohse}, T. and {Martin}, J.~M. and {Martineau-Huynh}, O. and {Marcowith}, A. and {Masterson}, C. and {McComb}, T.~J.~L. and {de Naurois}, M. and {Nolan}, S.~J. and {Noutsos}, A. and {Orford}, K.~J. and {Osborne}, J.~L. and {Ouchrif}, M. and {Panter}, M. and {Pelletier}, G. and {Pita}, S. and {P{\"u}hlhofer}, G. and {Punch}, M. and {Raubenheimer}, B.~C. and {Raue}, M. and {Raux}, J. and {Rayner}, S.~M. and {Reimer}, A. and {Reimer}, O. and {Ripken}, J. and {Rob}, L. and {Rolland}, L. and {Rowell}, G. and {Sahakian}, V. and {Saug{\'e}}, L. and {Schlenker}, S. and {Schlickeiser}, R. and {Schuster}, C. and {Schwanke}, U. and {Siewert}, M. and {Sol}, H. and {Spangler}, D. and {Steenkamp}, R. and {Stegmann}, C. and {Tavernet}, J. -P. and {Terrier}, R. and {Th{\'e}oret}, C.~G. and {Tluczykont}, M. and {Vasileiadis}, G. and {Venter}, C. and {Vincent}, P. and {V{\"o}lk}, H.~J. and {Wagner}, S.~J.},
  title         = {{The H.E.S.S. Survey of the Inner Galaxy in Very High Energy Gamma Rays}},
  journal       = {\apj},
  year          = 2006,
  month         = jan,
  volume        = {636},
  number        = {2},
  pages         = {777-797},
  doi           = {10.1086/498013},
  archiveprefix = {arXiv},
  eprint        = {astro-ph/0510397},
  primaryclass  = {astro-ph},
  adsurl        = {https://ui.adsabs.harvard.edu/abs/2006ApJ...636..777A}
}

@article{2006ApJ...652.1044R,
  author        = {{Rieger}, Frank M. and {Duffy}, Peter},
  title         = {{A Microscopic Analysis of Shear Acceleration}},
  journal       = {\apj},
  year          = 2006,
  month         = dec,
  volume        = {652},
  number        = {2},
  pages         = {1044-1049},
  doi           = {10.1086/508056},
  archiveprefix = {arXiv},
  eprint        = {astro-ph/0610187},
  primaryclass  = {astro-ph},
  adsurl        = {https://ui.adsabs.harvard.edu/abs/2006ApJ...652.1044R}
}

@article{2006Sci...314.1424A,
  author        = {{Aharonian}, F. and {Akhperjanian}, A.~G. and {Bazer-Bachi}, A.~R. and {Beilicke}, M. and {Benbow}, W. and {Berge}, D. and {Bernl{\"o}hr}, K. and {Boisson}, C. and {Bolz}, O. and {Borrel}, V. and {Braun}, I. and {Brown}, A.~M. and {B{\"u}hler}, R. and {B{\"u}sching}, I. and {Carrigan}, S. and {Chadwick}, P.~M. and {Chounet}, L. -M. and {Coignet}, G. and {Cornils}, R. and {Costamante}, L. and {Degrange}, B. and {Dickinson}, H.~J. and {Djannati-Ata{\"\i}}, A. and {Drury}, L. O'C. and {Dubus}, G. and {Egberts}, K. and {Emmanoulopoulos}, D. and {Espigat}, P. and {Feinstein}, F. and {Ferrero}, E. and {Fiasson}, A. and {Fontaine}, G. and {Funk}, Seb. and {Funk}, S. and {F{\"u}{\ss}ling}, M. and {Gallant}, Y.~A. and {Giebels}, B. and {Glicenstein}, J.~F. and {Goret}, P. and {Hadjichristidis}, C. and {Hauser}, D. and {Hauser}, M. and {Heinzelmann}, G. and {Henri}, G. and {Hermann}, G. and {Hinton}, J.~A. and {Hoffmann}, A. and {Hofmann}, W. and {Holleran}, M. and {Hoppe}, S. and {Horns}, D. and {Jacholkowska}, A. and {de Jager}, O.~C. and {Kendziorra}, E. and {Kerschhaggl}, M. and {Kh{\'e}lifi}, B. and {Komin}, Nu. and {Konopelko}, A. and {Kosack}, K. and {Lamanna}, G. and {Latham}, I.~J. and {Le Gallou}, R. and {Lemi{\`e}re}, A. and {Lemoine-Goumard}, M. and {Lenain}, J. -P. and {Lohse}, T. and {Martin}, J.~M. and {Martineau-Huynh}, O. and {Marcowith}, A. and {Masterson}, C. and {Maurin}, G. and {McComb}, T.~J.~L. and {Moulin}, E. and {de Naurois}, M. and {Nedbal}, D. and {Nolan}, S.~J. and {Noutsos}, A. and {Orford}, K.~J. and {Osborne}, J.~L. and {Ouchrif}, M. and {Panter}, M. and {Pelletier}, G. and {Pita}, S. and {P{\"u}hlhofer}, G. and {Punch}, M. and {Ranchon}, S. and {Raubenheimer}, B.~C. and {Raue}, M. and {Rayner}, S.~M. and {Reimer}, A. and {Ripken}, J. and {Rob}, L. and {Rolland}, L. and {Rosier-Lees}, S. and {Rowell}, G. and {Sahakian}, V. and {Santangelo}, A. and {Saug{\'e}}, L. and {Schlenker}, S. and {Schlickeiser}, R. and {Schr{\"o}der}, R. and {Schwanke}, U. and {Schwarzburg}, S. and {Schwemmer}, S. and {Shalchi}, A. and {Sol}, H. and {Spangler}, D. and {Spanier}, F. and {Steenkamp}, R. and {Stegmann}, C. and {Superina}, G. and {Tam}, P.~H. and {Tavernet}, J. -P. and {Terrier}, R. and {Tluczykont}, M. and {van Eldik}, C. and {Vasileiadis}, G. and {Venter}, C. and {Vialle}, J.~P. and {Vincent}, P. and {V{\"o}lk}, H.~J. and {Wagner}, S.~J. and {Ward}, M.},
  title         = {{Fast Variability of Tera-Electron Volt {\ensuremath{\gamma}} Rays from the Radio Galaxy M87}},
  journal       = {Science},
  year          = 2006,
  month         = dec,
  volume        = {314},
  number        = {5804},
  pages         = {1424-1427},
  doi           = {10.1126/science.1134408},
  archiveprefix = {arXiv},
  eprint        = {astro-ph/0612016},
  primaryclass  = {astro-ph},
  adsurl        = {https://ui.adsabs.harvard.edu/abs/2006Sci...314.1424A}
}

@article{2011ApJ...735..102K,
  author        = {{Kowal}, Grzegorz and {de Gouveia Dal Pino}, E.~M. and {Lazarian}, A.},
  title         = {{Magnetohydrodynamic Simulations of Reconnection and Particle Acceleration: Three-dimensional Effects}},
  journal       = {\apj},
  year          = 2011,
  month         = jul,
  volume        = {735},
  number        = {2},
  eid           = {102},
  pages         = {102},
  doi           = {10.1088/0004-637X/735/2/102},
  archiveprefix = {arXiv},
  eprint        = {1103.2984},
  primaryclass  = {astro-ph.HE},
  adsurl        = {https://ui.adsabs.harvard.edu/abs/2011ApJ...735..102K}
}

@article{2011ApJ...737L..40U,
  author        = {{Uzdensky}, Dmitri A. and {Cerutti}, Beno{\^\i}t and {Begelman}, Mitchell C.},
  title         = {{Reconnection-powered Linear Accelerator and Gamma-Ray Flares in the Crab Nebula}},
  journal       = {\apjl},
  year          = 2011,
  month         = aug,
  volume        = {737},
  number        = {2},
  eid           = {L40},
  pages         = {L40},
  doi           = {10.1088/2041-8205/737/2/L40},
  archiveprefix = {arXiv},
  eprint        = {1105.0942},
  primaryclass  = {astro-ph.HE},
  adsurl        = {https://ui.adsabs.harvard.edu/abs/2011ApJ...737L..40U}
}

@article{2011ApJ...743...47B,
  author   = {{Becker}, Peter A. and {Das}, Santabrata and {Le}, Truong},
  title    = {{Diffusive Particle Acceleration in Shocked, Viscous Accretion Disks: Green's Function Energy Distribution}},
  journal  = {\apj},
  year     = 2011,
  month    = dec,
  volume   = {743},
  number   = {1},
  eid      = {47},
  pages    = {47},
  doi      = {10.1088/0004-637X/743/1/47},
  adsurl   = {https://ui.adsabs.harvard.edu/abs/2011ApJ...743...47B}
}

@article{2012A&A...544A..96A,
  author        = {{Aleksi{\'c}}, J. and {Alvarez}, E.~A. and {Antonelli}, L.~A. and {Antoranz}, P. and {Asensio}, M. and {Backes}, M. and {Barrio}, J.~A. and {Bastieri}, D. and {Becerra Gonz{\'a}lez}, J. and {Bednarek}, W. and {Berdyugin}, A. and {Berger}, K. and {Bernardini}, E. and {Biland}, A. and {Blanch}, O. and {Bock}, R.~K. and {Boller}, A. and {Bonnoli}, G. and {Borla Tridon}, D. and {Braun}, I. and {Bretz}, T. and {Ca{\~n}ellas}, A. and {Carmona}, E. and {Carosi}, A. and {Colin}, P. and {Colombo}, E. and {Contreras}, J.~L. and {Cortina}, J. and {Cossio}, L. and {Covino}, S. and {Dazzi}, F. and {De Angelis}, A. and {De Caneva}, G. and {De Cea del Pozo}, E. and {De Lotto}, B. and {Delgado Mendez}, C. and {Diago Ortega}, A. and {Doert}, M. and {Dom{\'\i}nguez}, A. and {Dominis Prester}, D. and {Dorner}, D. and {Doro}, M. and {Elsaesser}, D. and {Ferenc}, D. and {Fonseca}, M.~V. and {Font}, L. and {Fruck}, C. and {Garc{\'\i}a L{\'o}pez}, R.~J. and {Garczarczyk}, M. and {Garrido}, D. and {Gaug}, M. and {Giavitto}, G. and {Godinovi{\'c}}, N. and {Hadasch}, D. and {H{\"a}fner}, D. and {Herrero}, A. and {Hildebrand}, D. and {H{\"o}hne-M{\"o}nch}, D. and {Hose}, J. and {Hrupec}, D. and {Huber}, B. and {Jogler}, T. and {Kellermann}, H. and {Klepser}, S. and {Kr{\"a}henb{\"u}hl}, T. and {Krause}, J. and {La Barbera}, A. and {Lelas}, D. and {Leonardo}, E. and {Lindfors}, E. and {Lombardi}, S. and {L{\'o}pez}, M. and {L{\'o}pez-Oramas}, A. and {Lorenz}, E. and {Makariev}, M. and {Maneva}, G. and {Mankuzhiyil}, N. and {Mannheim}, K. and {Maraschi}, L. and {Mariotti}, M. and {Mart{\'\i}nez}, M. and {Mazin}, D. and {Meucci}, M. and {Miranda}, J.~M. and {Mirzoyan}, R. and {Miyamoto}, H. and {Mold{\'o}n}, J. and {Moralejo}, A. and {Munar-Adrover}, P. and {Nieto}, D. and {Nilsson}, K. and {Orito}, R. and {Oya}, I. and {Paneque}, D. and {Paoletti}, R. and {Pardo}, S. and {Paredes}, J.~M. and {Partini}, S. and {Pasanen}, M. and {Pauss}, F. and {Perez-Torres}, M.~A. and {Persic}, M. and {Peruzzo}, L. and {Pilia}, M. and {Pochon}, J. and {Prada}, F. and {Prada Moroni}, P.~G. and {Prandini}, E. and {Puljak}, I. and {Reichardt}, I. and {Reinthal}, R. and {Rhode}, W. and {Rib{\'o}}, M. and {Rico}, J. and {R{\"u}gamer}, S. and {Saggion}, A. and {Saito}, K. and {Saito}, T.~Y. and {Salvati}, M. and {Satalecka}, K. and {Scalzotto}, V. and {Scapin}, V. and {Schultz}, C. and {Schweizer}, T. and {Shayduk}, M. and {Shore}, S.~N. and {Sillanp{\"a}{\"a}}, A. and {Sitarek}, J. and {Snidaric}, I. and {Sobczynska}, D. and {Spanier}, F. and {Spiro}, S. and {Stamatescu}, V. and {Stamerra}, A. and {Steinke}, B. and {Storz}, J. and {Strah}, N. and {Suri{\'c}}, T. and {Takalo}, L. and {Takami}, H. and {Tavecchio}, F. and {Temnikov}, P. and {Terzi{\'c}}, T. and {Tescaro}, D. and {Teshima}, M. and {Tibolla}, O. and {Torres}, D.~F. and {Treves}, A. and {Uellenbeck}, M. and {Vankov}, H. and {Vogler}, P. and {Wagner}, R.~M. and {Weitzel}, Q. and {Zabalza}, V. and {Zandanel}, F. and {Zanin}, R. and {Ghisellini}, G.},
  title         = {{MAGIC observations of the giant radio galaxy M 87 in a low-emission state between 2005 and 2007}},
  journal       = {\aap},
  year          = 2012,
  month         = aug,
  volume        = {544},
  eid           = {A96},
  pages         = {A96},
  doi           = {10.1051/0004-6361/201117827},
  archiveprefix = {arXiv},
  eprint        = {1207.2147},
  primaryclass  = {astro-ph.HE},
  adsurl        = {https://ui.adsabs.harvard.edu/abs/2012A&A...544A..96A}
}

@article{2012ApJ...746..141A,
  author        = {{Aliu}, E. and {Arlen}, T. and {Aune}, T. and {Beilicke}, M. and {Benbow}, W. and {Bouvier}, A. and {Bradbury}, S.~M. and {Buckley}, J.~H. and {Bugaev}, V. and {Byrum}, K. and {Cannon}, A. and {Cesarini}, A. and {Ciupik}, L. and {Collins-Hughes}, E. and {Connolly}, M.~P. and {Cui}, W. and {Dickherber}, R. and {Duke}, C. and {Errando}, M. and {Falcone}, A. and {Finley}, J.~P. and {Finnegan}, G. and {Fortson}, L. and {Furniss}, A. and {Galante}, N. and {Gall}, D. and {Godambe}, S. and {Griffin}, S. and {Grube}, J. and {Guenette}, R. and {Gyuk}, G. and {Hanna}, D. and {Holder}, J. and {Huan}, H. and {Hughes}, G. and {Hui}, C.~M. and {Humensky}, T.~B. and {Imran}, A. and {Kaaret}, P. and {Karlsson}, N. and {Kertzman}, M. and {Kieda}, D. and {Krawczynski}, H. and {Krennrich}, F. and {Lang}, M.~J. and {LeBohec}, S. and {Madhavan}, A.~S. and {Maier}, G. and {Majumdar}, P. and {McArthur}, S. and {McCann}, A. and {Moriarty}, P. and {Mukherjee}, R. and {Nu{\~n}ez}, P.~D. and {Ong}, R.~A. and {Orr}, M. and {Otte}, A.~N. and {Park}, N. and {Perkins}, J.~S. and {Pichel}, A. and {Pohl}, M. and {Prokoph}, H. and {Quinn}, J. and {Ragan}, K. and {Reyes}, L.~C. and {Reynolds}, P.~T. and {Roache}, E. and {Rose}, H.~J. and {Ruppel}, J. and {Saxon}, D.~B. and {Schroedter}, M. and {Sembroski}, G.~H. and {{\c{S}}ent{\"u}rk}, G.~D. and {Skole}, C. and {Staszak}, D. and {Te{\v{s}}i{\'c}}, G. and {Theiling}, M. and {Thibadeau}, S. and {Tsurusaki}, K. and {Tyler}, J. and {Varlotta}, A. and {Vassiliev}, V.~V. and {Vincent}, S. and {Vivier}, M. and {Wakely}, S.~P. and {Ward}, J.~E. and {Weekes}, T.~C. and {Weinstein}, A. and {Weisgarber}, T. and {Williams}, D.~A. and {Zitzer}, B.},
  title         = {{VERITAS Observations of Day-scale Flaring of M 87 in 2010 April}},
  journal       = {\apj},
  year          = 2012,
  month         = feb,
  volume        = {746},
  number        = {2},
  eid           = {141},
  pages         = {141},
  doi           = {10.1088/0004-637X/746/2/141},
  archiveprefix = {arXiv},
  eprint        = {1112.4518},
  primaryclass  = {astro-ph.CO},
  adsurl        = {https://ui.adsabs.harvard.edu/abs/2012ApJ...746..141A}
}

@article{2012ApJ...755...50R,
  author        = {{Riquelme}, Mario A. and {Quataert}, Eliot and {Sharma}, Prateek and {Spitkovsky}, Anatoly},
  title         = {{Local Two-dimensional Particle-in-cell Simulations of the Collisionless Magnetorotational Instability}},
  journal       = {\apj},
  year          = 2012,
  month         = aug,
  volume        = {755},
  number        = {1},
  eid           = {50},
  pages         = {50},
  doi           = {10.1088/0004-637X/755/1/50},
  archiveprefix = {arXiv},
  eprint        = {1201.6407},
  primaryclass  = {astro-ph.HE},
  adsurl        = {https://ui.adsabs.harvard.edu/abs/2012ApJ...755...50R}
}

@article{2012arXiv1203.6851M,
  author        = {{Montesinos}, Matias},
  title         = {{Review: Accretion Disk Theory}},
  journal       = {arXiv e-prints},
  year          = 2012,
  month         = mar,
  eid           = {arXiv:1203.6851},
  pages         = {arXiv:1203.6851},
  doi           = {10.48550/arXiv.1203.6851},
  archiveprefix = {arXiv},
  eprint        = {1203.6851},
  primaryclass  = {astro-ph.HE},
  adsurl        = {https://ui.adsabs.harvard.edu/abs/2012arXiv1203.6851M}
}

@article{2012PhRvL.108x1102K,
  author        = {{Kowal}, Grzegorz and {de Gouveia Dal Pino}, Elisabete M. and {Lazarian}, A.},
  title         = {{Particle Acceleration in Turbulence and Weakly Stochastic Reconnection}},
  journal       = {\prl},
  year          = 2012,
  month         = jun,
  volume        = {108},
  number        = {24},
  eid           = {241102},
  pages         = {241102},
  doi           = {10.1103/PhysRevLett.108.241102},
  archiveprefix = {arXiv},
  eprint        = {1202.5256},
  primaryclass  = {astro-ph.HE},
  adsurl        = {https://ui.adsabs.harvard.edu/abs/2012PhRvL.108x1102K}
}

@article{2012SSRv..173..557L,
  author        = {{Lazarian}, A. and {Vlahos}, L. and {Kowal}, G. and {Yan}, H. and {Beresnyak}, A. and {de Gouveia Dal Pino}, E.~M.},
  title         = {{Turbulence, Magnetic Reconnection in Turbulent Fluids and Energetic Particle Acceleration}},
  journal       = {\ssr},
  year          = 2012,
  month         = nov,
  volume        = {173},
  number        = {1-4},
  pages         = {557-622},
  doi           = {10.1007/s11214-012-9936-7},
  archiveprefix = {arXiv},
  eprint        = {1211.0008},
  primaryclass  = {astro-ph.SR},
  adsurl        = {https://ui.adsabs.harvard.edu/abs/2012SSRv..173..557L}
}

@article{2013ApJ...767...30B,
  author        = {{Bai}, Xue-Ning and {Stone}, James M.},
  title         = {{Local Study of Accretion Disks with a Strong Vertical Magnetic Field: Magnetorotational Instability and Disk Outflow}},
  journal       = {\apj},
  year          = 2013,
  month         = apr,
  volume        = {767},
  number        = {1},
  eid           = {30},
  pages         = {30},
  doi           = {10.1088/0004-637X/767/1/30},
  archiveprefix = {arXiv},
  eprint        = {1210.6661},
  primaryclass  = {astro-ph.HE},
  adsurl        = {https://ui.adsabs.harvard.edu/abs/2013ApJ...767...30B}
}

@article{2013ApJ...767L..16O,
  author        = {{Ohira}, Yutaka},
  title         = {{Turbulent Shear Acceleration}},
  journal       = {\apjl},
  year          = 2013,
  month         = apr,
  volume        = {767},
  number        = {1},
  eid           = {L16},
  pages         = {L16},
  doi           = {10.1088/2041-8205/767/1/L16},
  archiveprefix = {arXiv},
  eprint        = {1301.6573},
  primaryclass  = {astro-ph.HE},
  adsurl        = {https://ui.adsabs.harvard.edu/abs/2013ApJ...767L..16O}
}

@article{2013ApJ...773..118H,
  author        = {{Hoshino}, Masahiro},
  title         = {{Particle Acceleration during Magnetorotational Instability in a Collisionless Accretion Disk}},
  journal       = {\apj},
  year          = 2013,
  month         = aug,
  volume        = {773},
  number        = {2},
  eid           = {118},
  pages         = {118},
  doi           = {10.1088/0004-637X/773/2/118},
  archiveprefix = {arXiv},
  eprint        = {1306.6720},
  primaryclass  = {astro-ph.HE},
  adsurl        = {https://ui.adsabs.harvard.edu/abs/2013ApJ...773..118H}
}

@article{2013LRR....16....1A,
  author        = {{Abramowicz}, Marek A. and {Fragile}, P. Chris},
  title         = {{Foundations of Black Hole Accretion Disk Theory}},
  journal       = {Living Reviews in Relativity},
  year          = 2013,
  month         = dec,
  volume        = {16},
  number        = {1},
  eid           = {1},
  pages         = {1},
  doi           = {10.12942/lrr-2013-1},
  archiveprefix = {arXiv},
  eprint        = {1104.5499},
  primaryclass  = {astro-ph.HE},
  adsurl        = {https://ui.adsabs.harvard.edu/abs/2013LRR....16....1A}
}

@article{2014ApJ...783L..21S,
  author        = {{Sironi}, Lorenzo and {Spitkovsky}, Anatoly},
  title         = {{Relativistic Reconnection: An Efficient Source of Non-thermal Particles}},
  journal       = {\apjl},
  year          = 2014,
  month         = mar,
  volume        = {783},
  number        = {1},
  eid           = {L21},
  pages         = {L21},
  doi           = {10.1088/2041-8205/783/1/L21},
  archiveprefix = {arXiv},
  eprint        = {1401.5471},
  primaryclass  = {astro-ph.HE},
  adsurl        = {https://ui.adsabs.harvard.edu/abs/2014ApJ...783L..21S}
}

@inproceedings{2014ASPC..488....8D,
  author        = {{de Gouveia Dal Pino}, E.~M. and {Kowal}, G. and {Lazarian}, A.},
  title         = {{Fermi Acceleration in Magnetic Reconnection Sites}},
  booktitle     = {8th International Conference of Numerical Modeling of Space Plasma Flows (ASTRONUM 2013)},
  year          = 2014,
  editor        = {{Pogorelov}, N.~V. and {Audit}, E. and {Zank}, G.~P.},
  series        = {Astronomical Society of the Pacific Conference Series},
  volume        = {488},
  month         = sep,
  pages         = {8},
  doi           = {10.48550/arXiv.1401.4941},
  archiveprefix = {arXiv},
  eprint        = {1401.4941},
  primaryclass  = {astro-ph.HE},
  adsurl        = {https://ui.adsabs.harvard.edu/abs/2014ASPC..488....8D}
}

@article{2015PhRvL.114f1101H,
  author        = {{Hoshino}, Masahiro},
  title         = {{Angular Momentum Transport and Particle Acceleration During Magnetorotational Instability in a Kinetic Accretion Disk}},
  journal       = {\prl},
  year          = 2015,
  month         = feb,
  volume        = {114},
  number        = {6},
  eid           = {061101},
  pages         = {061101},
  doi           = {10.1103/PhysRevLett.114.061101},
  archiveprefix = {arXiv},
  eprint        = {1502.02452},
  primaryclass  = {astro-ph.HE},
  adsurl        = {https://ui.adsabs.harvard.edu/abs/2015PhRvL.114f1101H}
}

@article{2016ApJ...816L...8W,
  author        = {{Werner}, G.~R. and {Uzdensky}, D.~A. and {Cerutti}, B. and {Nalewajko}, K. and {Begelman}, M.~C.},
  title         = {{The Extent of Power-law Energy Spectra in Collisionless Relativistic Magnetic Reconnection in Pair Plasmas}},
  journal       = {\apjl},
  year          = 2016,
  month         = jan,
  volume        = {816},
  number        = {1},
  eid           = {L8},
  pages         = {L8},
  doi           = {10.3847/2041-8205/816/1/L8},
  archiveprefix = {arXiv},
  eprint        = {1409.8262},
  primaryclass  = {astro-ph.HE},
  adsurl        = {https://ui.adsabs.harvard.edu/abs/2016ApJ...816L...8W}
}

@article{2016ApJ...822...88K,
  author        = {{Kimura}, Shigeo S. and {Toma}, Kenji and {Suzuki}, Takeru K. and {Inutsuka}, Shu-ichiro},
  title         = {{Stochastic Particle Acceleration in Turbulence Generated by Magnetorotational Instability}},
  journal       = {\apj},
  year          = 2016,
  month         = may,
  volume        = {822},
  number        = {2},
  eid           = {88},
  pages         = {88},
  doi           = {10.3847/0004-637X/822/2/88},
  archiveprefix = {arXiv},
  eprint        = {1602.07773},
  primaryclass  = {astro-ph.HE},
  adsurl        = {https://ui.adsabs.harvard.edu/abs/2016ApJ...822...88K}
}

@article{2016MNRAS.457..857S,
  author        = {{Salvesen}, Greg and {Simon}, Jacob B. and {Armitage}, Philip J. and {Begelman}, Mitchell C.},
  title         = {{Accretion disc dynamo activity in local simulations spanning weak-to-strong net vertical magnetic flux regimes}},
  journal       = {\mnras},
  year          = 2016,
  month         = mar,
  volume        = {457},
  number        = {1},
  pages         = {857-874},
  doi           = {10.1093/mnras/stw029},
  archiveprefix = {arXiv},
  eprint        = {1511.06368},
  primaryclass  = {astro-ph.HE},
  adsurl        = {https://ui.adsabs.harvard.edu/abs/2016MNRAS.457..857S}
}

@article{2016MNRAS.463.4331D,
  author        = {{del Valle}, Maria V. and {de Gouveia Dal Pino}, E.~M. and {Kowal}, G.},
  title         = {{Properties of the first-order Fermi acceleration in fast magnetic reconnection driven by turbulence in collisional magnetohydrodynamical flows}},
  journal       = {\mnras},
  year          = 2016,
  month         = dec,
  volume        = {463},
  number        = {4},
  pages         = {4331-4343},
  doi           = {10.1093/mnras/stw2276},
  archiveprefix = {arXiv},
  eprint        = {1609.08598},
  primaryclass  = {astro-ph.HE},
  adsurl        = {https://ui.adsabs.harvard.edu/abs/2016MNRAS.463.4331D}
}

@article{2016Natur.531..476H,
  author        = {{H.~E.~S.~S. Collaboration} and {Abramowski}, A. and {Aharonian}, F. and {Benkhali}, F. Ait and {Akhperjanian}, A.~G. and {Ang{\"u}ner}, E.~O. and {Backes}, M. and {Balzer}, A. and {Becherini}, Y. and {Tjus}, J. Becker and {Berge}, D. and {Bernhard}, S. and {Bernl{\"o}hr}, K. and {Birsin}, E. and {Blackwell}, R. and {B{\"o}ttcher}, M. and {Boisson}, C. and {Bolmont}, J. and {Bordas}, P. and {Bregeon}, J. and {Brun}, F. and {Brun}, P. and {Bryan}, M. and {Bulik}, T. and {Carr}, J. and {Casanova}, S. and {Chakraborty}, N. and {Chalme-Calvet}, R. and {Chaves}, R.~C.~G. and {Chen}, A. and {Chr{\'e}tien}, M. and {Colafrancesco}, S. and {Cologna}, G. and {Conrad}, J. and {Couturier}, C. and {Cui}, Y. and {Davids}, I.~D. and {Degrange}, B. and {Deil}, C. and {Dewilt}, P. and {Djannati-Ata{\"\i}}, A. and {Domainko}, W. and {Donath}, A. and {Drury}, L. O'C. and {Dubus}, G. and {Dutson}, K. and {Dyks}, J. and {Dyrda}, M. and {Edwards}, T. and {Egberts}, K. and {Eger}, P. and {Ernenwein}, J. -P. and {Espigat}, P. and {Farnier}, C. and {Fegan}, S. and {Feinstein}, F. and {Fernandes}, M.~V. and {Fernandez}, D. and {Fiasson}, A. and {Fontaine}, G. and {F{\"o}rster}, A. and {F{\"u}{\ss}ling}, M. and {Gabici}, S. and {Gajdus}, M. and {Gallant}, Y.~A. and {Garrigoux}, T. and {Giavitto}, G. and {Giebels}, B. and {Glicenstein}, J.~F. and {Gottschall}, D. and {Goyal}, A. and {Grondin}, M. -H. and {Grudzi{\'n}ska}, M. and {Hadasch}, D. and {H{\"a}ffner}, S. and {Hahn}, J. and {Hawkes}, J. and {Heinzelmann}, G. and {Henri}, G. and {Hermann}, G. and {Hervet}, O. and {Hillert}, A. and {Hinton}, J.~A. and {Hofmann}, W. and {Hofverberg}, P. and {Hoischen}, C. and {Holler}, M. and {Horns}, D. and {Ivascenko}, A. and {Jacholkowska}, A. and {Jamrozy}, M. and {Janiak}, M. and {Jankowsky}, F. and {Jung-Richardt}, I. and {Kastendieck}, M.~A. and {Katarzy{\'n}ski}, K. and {Katz}, U. and {Kerszberg}, D. and {Kh{\'e}lifi}, B. and {Kieffer}, M. and {Klepser}, S. and {Klochkov}, D. and {Klu{\'z}niak}, W. and {Kolitzus}, D. and {Komin}, Nu. and {Kosack}, K. and {Krakau}, S. and {Krayzel}, F. and {Kr{\"u}ger}, P.~P. and {Laffon}, H. and {Lamanna}, G. and {Lau}, J. and {Lefaucheur}, J. and {Lefranc}, V. and {Lemi{\'e}re}, A. and {Lemoine-Goumard}, M. and {Lenain}, J. -P. and {Lohse}, T. and {Lopatin}, A. and {Lu}, C. -C. and {Lui}, R. and {Marandon}, V. and {Marcowith}, A. and {Mariaud}, C. and {Marx}, R. and {Maurin}, G. and {Maxted}, N. and {Mayer}, M. and {Meintjes}, P.~J. and {Menzler}, U. and {Meyer}, M. and {Mitchell}, A.~M.~W. and {Moderski}, R. and {Mohamed}, M. and {Mor{\r{a}}}, K. and {Moulin}, E. and {Murach}, T. and {de Naurois}, M. and {Niemiec}, J. and {Oakes}, L. and {Odaka}, H. and {{\"O}ttl}, S. and {Ohm}, S. and {Opitz}, B. and {Ostrowski}, M. and {Oya}, I. and {Panter}, M. and {Parsons}, R.~D. and {Arribas}, M. Paz and {Pekeur}, N.~W. and {Pelletier}, G. and {Petrucci}, P. -O. and {Peyaud}, B. and {Pita}, S. and {Poon}, H. and {Prokoph}, H. and {P{\"u}hlhofer}, G. and {Punch}, M. and {Quirrenbach}, A. and {Raab}, S. and {Reichardt}, I. and {Reimer}, A. and {Reimer}, O. and {Renaud}, M. and {de Los Reyes}, R. and {Rieger}, F. and {Romoli}, C. and {Rosier-Lees}, S. and {Rowell}, G. and {Rudak}, B. and {Rulten}, C.~B. and {Sahakian}, V. and {Salek}, D. and {Sanchez}, D.~A. and {Santangelo}, A. and {Sasaki}, M. and {Schlickeiser}, R. and {Sch{\"u}ssler}, F. and {Schulz}, A. and {Schwanke}, U. and {Schwemmer}, S. and {Seyffert}, A.~S. and {Simoni}, R. and {Sol}, H. and {Spanier}, F. and {Spengler}, G. and {Spies}, F. and {Stawarz}, {\L}. and {Steenkamp}, R. and {Stegmann}, C. and {Stinzing}, F. and {Stycz}, K. and {Sushch}, I. and {Tavernet}, J. -P. and {Tavernier}, T. and {Taylor}, A.~M. and {Terrier}, R. and {Tluczykont}, M. and {Trichard}, C. and {Tuffs}, R. and {Valerius}, K. and {van der Walt}, J. and {van Eldik}, C. and {van Soelen}, B. and {Vasileiadis}, G. and {Veh}, J. and {Venter}, C. and {Viana}, A. and {Vincent}, P. and {Vink}, J. and {Voisin}, F. and {V{\"o}lk}, H.~J. and {Vuillaume}, T. and {Wagner}, S.~J. and {Wagner}, P. and {Wagner}, R.~M. and {Weidinger}, M. and {Weitzel}, Q. and {White}, R. and {Wierzcholska}, A. and {Willmann}, P. and {W{\"o}rnlein}, A. and {Wouters}, D. and {Yang}, R. and {Zabalza}, V. and {Zaborov}, D. and {Zacharias}, M. and {Zdziarski}, A.~A. and {Zech}, A. and {Zefi}, F. and {{\.Z}ywucka}, N.},
  title         = {{Acceleration of petaelectronvolt protons in the Galactic Centre}},
  journal       = {\nat},
  year          = 2016,
  month         = mar,
  volume        = {531},
  number        = {7595},
  pages         = {476-479},
  doi           = {10.1038/nature17147},
  archiveprefix = {arXiv},
  eprint        = {1603.07730},
  primaryclass  = {astro-ph.HE},
  adsurl        = {https://ui.adsabs.harvard.edu/abs/2016Natur.531..476H}
}

@article{2017MNRAS.465.1409L,
  author        = {{Lee}, Jason P. and {Becker}, Peter A.},
  title         = {{A two-fluid model for black-hole accretion flows: particle acceleration and disc structure}},
  journal       = {\mnras},
  year          = 2017,
  month         = feb,
  volume        = {465},
  number        = {2},
  pages         = {1409-1442},
  doi           = {10.1093/mnras/stw2787},
  archiveprefix = {arXiv},
  eprint        = {2002.06132},
  primaryclass  = {astro-ph.HE},
  adsurl        = {https://ui.adsabs.harvard.edu/abs/2017MNRAS.465.1409L}
}

@article{2017MNRAS.468.2447P,
  author        = {{Ponti}, G. and {George}, E. and {Scaringi}, S. and {Zhang}, S. and {Jin}, C. and {Dexter}, J. and {Terrier}, R. and {Clavel}, M. and {Degenaar}, N. and {Eisenhauer}, F. and {Genzel}, R. and {Gillessen}, S. and {Goldwurm}, A. and {Habibi}, M. and {Haggard}, D. and {Hailey}, C. and {Harrison}, F. and {Merloni}, A. and {Mori}, K. and {Nandra}, K. and {Ott}, T. and {Pfuhl}, O. and {Plewa}, P.~M. and {Waisberg}, I.},
  title         = {{A powerful flare from Sgr A* confirms the synchrotron nature of the X-ray emission}},
  journal       = {\mnras},
  year          = 2017,
  month         = jun,
  volume        = {468},
  number        = {2},
  pages         = {2447-2468},
  doi           = {10.1093/mnras/stx596},
  archiveprefix = {arXiv},
  eprint        = {1703.03410},
  primaryclass  = {astro-ph.HE},
  adsurl        = {https://ui.adsabs.harvard.edu/abs/2017MNRAS.468.2447P}
}

@article{2018A&A...612A...9H,
  author        = {{H.~E.~S.~S. Collaboration} and {Abdalla}, H. and {Abramowski}, A. and {Aharonian}, F. and {Ait Benkhali}, F. and {Akhperjanian}, A.~G. and {Andersson}, T. and {Ang{\"u}ner}, E.~O. and {Arakawa}, M. and {Arrieta}, M. and {Aubert}, P. and {Backes}, M. and {Balzer}, A. and {Barnard}, M. and {Becherini}, Y. and {Becker Tjus}, J. and {Berge}, D. and {Bernhard}, S. and {Bernl{\"o}hr}, K. and {Blackwell}, R. and {B{\"o}ttcher}, M. and {Boisson}, C. and {Bolmont}, J. and {Bonnefoy}, S. and {Bordas}, P. and {Bregeon}, J. and {Brun}, F. and {Brun}, P. and {Bryan}, M. and {B{\"u}chele}, M. and {Bulik}, T. and {Capasso}, M. and {Carr}, J. and {Casanova}, S. and {Cerruti}, M. and {Chakraborty}, N. and {Chaves}, R.~C.~G. and {Chen}, A. and {Chevalier}, J. and {Coffaro}, M. and {Colafrancesco}, S. and {Cologna}, G. and {Condon}, B. and {Conrad}, J. and {Cui}, Y. and {Davids}, I.~D. and {Decock}, J. and {Degrange}, B. and {Deil}, C. and {Devin}, J. and {deWilt}, P. and {Dirson}, L. and {Djannati-Ata{\"\i}}, A. and {Domainko}, W. and {Donath}, A. and {Drury}, L.~O. 'C. and {Dutson}, K. and {Dyks}, J. and {Edwards}, T. and {Egberts}, K. and {Eger}, P. and {Ernenwein}, J. -P. and {Eschbach}, S. and {Farnier}, C. and {Fegan}, S. and {Fernandes}, M.~V. and {Fiasson}, A. and {Fontaine}, G. and {F{\"o}rster}, A. and {Funk}, S. and {F{\"u}{\ss}ling}, M. and {Gabici}, S. and {Gallant}, Y.~A. and {Garrigoux}, T. and {Giavitto}, G. and {Giebels}, B. and {Glicenstein}, J.~F. and {Gottschall}, D. and {Goyal}, A. and {Grondin}, M. -H. and {Hahn}, J. and {Haupt}, M. and {Hawkes}, J. and {Heinzelmann}, G. and {Henri}, G. and {Hermann}, G. and {Hinton}, J.~A. and {Hofmann}, W. and {Hoischen}, C. and {Holch}, T.~L. and {Holler}, M. and {Horns}, D. and {Ivascenko}, A. and {Iwasaki}, H. and {Jacholkowska}, A. and {Jamrozy}, M. and {Janiak}, M. and {Jankowsky}, D. and {Jankowsky}, F. and {Jingo}, M. and {Jogler}, T. and {Jouvin}, L. and {Jung-Richardt}, I. and {Kastendieck}, M.~A. and {Katarzy{\'n}ski}, K. and {Katsuragawa}, M. and {Katz}, U. and {Kerszberg}, D. and {Khangulyan}, D. and {Kh{\'e}lifi}, B. and {King}, J. and {Klepser}, S. and {Klochkov}, D. and {Klu{\'z}niak}, W. and {Kolitzus}, D. and {Komin}, Nu. and {Kosack}, K. and {Krakau}, S. and {Kraus}, M. and {Kr{\"u}ger}, P.~P. and {Laffon}, H. and {Lamanna}, G. and {Lau}, J. and {Lees}, J. -P. and {Lefaucheur}, J. and {Lefranc}, V. and {Lemi{\`e}re}, A. and {Lemoine-Goumard}, M. and {Lenain}, J. -P. and {Leser}, E. and {Lohse}, T. and {Lorentz}, M. and {Liu}, R. and {L{\'o}pez-Coto}, R. and {Lypova}, I. and {Marandon}, V. and {Marcowith}, A. and {Mariaud}, C. and {Marx}, R. and {Maurin}, G. and {Maxted}, N. and {Mayer}, M. and {Meintjes}, P.~J. and {Meyer}, M. and {Mitchell}, A.~M.~W. and {Moderski}, R. and {Mohamed}, M. and {Mohrmann}, L. and {Mor{\r{a}}}, K. and {Moulin}, E. and {Murach}, T. and {Nakashima}, S. and {de Naurois}, M. and {Niederwanger}, F. and {Niemiec}, J. and {Oakes}, L. and {O'Brien}, P. and {Odaka}, H. and {Ohm}, S. and {Ostrowski}, M. and {Oya}, I. and {Padovani}, M. and {Panter}, M. and {Parsons}, R.~D. and {Pekeur}, N.~W. and {Pelletier}, G. and {Perennes}, C. and {Petrucci}, P. -O. and {Peyaud}, B. and {Piel}, Q. and {Pita}, S. and {Poon}, H. and {Prokhorov}, D. and {Prokoph}, H. and {P{\"u}hlhofer}, G. and {Punch}, M. and {Quirrenbach}, A. and {Raab}, S. and {Rauth}, R. and {Reimer}, A. and {Reimer}, O. and {Renaud}, M. and {de los Reyes}, R. and {Richter}, S. and {Rieger}, F. and {Romoli}, C. and {Rowell}, G. and {Rudak}, B. and {Rulten}, C.~B. and {Sahakian}, V. and {Saito}, S. and {Salek}, D. and {Sanchez}, D.~A. and {Santangelo}, A. and {Sasaki}, M. and {Schlickeiser}, R. and {Sch{\"u}ssler}, F. and {Schulz}, A. and {Schwanke}, U. and {Schwemmer}, S.},
  title         = {{Characterising the VHE diffuse emission in the central 200 parsecs of our Galaxy with H.E.S.S.}},
  journal       = {\aap},
  year          = 2018,
  month         = apr,
  volume        = {612},
  eid           = {A9},
  pages         = {A9},
  doi           = {10.1051/0004-6361/201730824},
  archiveprefix = {arXiv},
  eprint        = {1706.04535},
  primaryclass  = {astro-ph.HE},
  adsurl        = {https://ui.adsabs.harvard.edu/abs/2018A&A...612A...9H}
}

@article{2018A&A...618L..10G,
  author        = {{Gravity Collaboration} and {Abuter}, R. and {Amorim}, A. and {Baub{\"o}ck}, M. and {Berger}, J.~P. and {Bonnet}, H. and {Brandner}, W. and {Cl{\'e}net}, Y. and {Coud{\'e} Du Foresto}, V. and {de Zeeuw}, P.~T. and {Deen}, C. and {Dexter}, J. and {Duvert}, G. and {Eckart}, A. and {Eisenhauer}, F. and {F{\"o}rster Schreiber}, N.~M. and {Garcia}, P. and {Gao}, F. and {Gendron}, E. and {Genzel}, R. and {Gillessen}, S. and {Guajardo}, P. and {Habibi}, M. and {Haubois}, X. and {Henning}, Th. and {Hippler}, S. and {Horrobin}, M. and {Huber}, A. and {Jim{\'e}nez-Rosales}, A. and {Jocou}, L. and {Kervella}, P. and {Lacour}, S. and {Lapeyr{\`e}re}, V. and {Lazareff}, B. and {Le Bouquin}, J. -B. and {L{\'e}na}, P. and {Lippa}, M. and {Ott}, T. and {Panduro}, J. and {Paumard}, T. and {Perraut}, K. and {Perrin}, G. and {Pfuhl}, O. and {Plewa}, P.~M. and {Rabien}, S. and {Rodr{\'\i}guez-Coira}, G. and {Rousset}, G. and {Sternberg}, A. and {Straub}, O. and {Straubmeier}, C. and {Sturm}, E. and {Tacconi}, L.~J. and {Vincent}, F. and {von Fellenberg}, S. and {Waisberg}, I. and {Widmann}, F. and {Wieprecht}, E. and {Wiezorrek}, E. and {Woillez}, J. and {Yazici}, S.},
  title         = {{Detection of orbital motions near the last stable circular orbit of the massive black hole SgrA*}},
  journal       = {\aap},
  year          = 2018,
  month         = oct,
  volume        = {618},
  eid           = {L10},
  pages         = {L10},
  doi           = {10.1051/0004-6361/201834294},
  archiveprefix = {arXiv},
  eprint        = {1810.12641},
  primaryclass  = {astro-ph.GA},
  adsurl        = {https://ui.adsabs.harvard.edu/abs/2018A&A...618L..10G}
}

@article{2018ApJ...859..149I,
  author        = {{Inchingolo}, Giannandrea and {Grismayer}, Thomas and {Loureiro}, Nuno F. and {Fonseca}, Ricardo A. and {Silva}, Luis O.},
  title         = {{Fully Kinetic Large-scale Simulations of the Collisionless Magnetorotational Instability}},
  journal       = {\apj},
  year          = 2018,
  month         = jun,
  volume        = {859},
  number        = {2},
  eid           = {149},
  pages         = {149},
  doi           = {10.3847/1538-4357/aac0f2},
  archiveprefix = {arXiv},
  eprint        = {1801.08657},
  primaryclass  = {astro-ph.HE},
  adsurl        = {https://ui.adsabs.harvard.edu/abs/2018ApJ...859..149I}
}

@ARTICLE{2018ApJ...864..126R,
       author = {{Ryan}, Benjamin R. and {Ressler}, Sean M. and {Dolence}, Joshua C. and {Gammie}, Charles and {Quataert}, Eliot},
        title = "{Two-temperature GRRMHD Simulations of M87}",
      journal = {\apj},
         year = 2018,
        month = sep,
       volume = {864},
       number = {2},
          eid = {126},
        pages = {126},
          doi = {10.3847/1538-4357/aad73a},
archivePrefix = {arXiv},
       eprint = {1808.01958},
 primaryClass = {astro-ph.HE},
       adsurl = {https://ui.adsabs.harvard.edu/abs/2018ApJ...864..126R}
}

@article{2018MNRAS.481.5687P,
  author        = {{Petropoulou}, Maria and {Sironi}, Lorenzo},
  title         = {{The steady growth of the high-energy spectral cut-off in relativistic magnetic reconnection}},
  journal       = {\mnras},
  year          = 2018,
  month         = dec,
  volume        = {481},
  number        = {4},
  pages         = {5687-5701},
  doi           = {10.1093/mnras/sty2702},
  archiveprefix = {arXiv},
  eprint        = {1808.00966},
  primaryclass  = {astro-ph.HE},
  adsurl        = {https://ui.adsabs.harvard.edu/abs/2018MNRAS.481.5687P}
}

@article{2019MNRAS.485..163K,
  author        = {{Kimura}, Shigeo S. and {Tomida}, Kengo and {Murase}, Kohta},
  title         = {{Acceleration and escape processes of high-energy particles in turbulence inside hot accretion flows}},
  journal       = {\mnras},
  year          = 2019,
  month         = may,
  volume        = {485},
  number        = {1},
  pages         = {163-178},
  doi           = {10.1093/mnras/stz329},
  archiveprefix = {arXiv},
  eprint        = {1812.03901},
  primaryclass  = {astro-ph.HE},
  adsurl        = {https://ui.adsabs.harvard.edu/abs/2019MNRAS.485..163K}
}

@article{2019PhRvD..99h3006L,
  author        = {{Lemoine}, Martin},
  title         = {{Generalized Fermi acceleration}},
  journal       = {\prd},
  year          = 2019,
  month         = apr,
  volume        = {99},
  number        = {8},
  eid           = {083006},
  pages         = {083006},
  doi           = {10.1103/PhysRevD.99.083006},
  archiveprefix = {arXiv},
  eprint        = {1903.05917},
  primaryclass  = {astro-ph.HE},
  adsurl        = {https://ui.adsabs.harvard.edu/abs/2019PhRvD..99h3006L}
}

@article{2020ApJ...899..151K,
  author        = {{Kilian}, Patrick and {Li}, Xiaocan and {Guo}, Fan and {Li}, Hui},
  title         = {{Exploring the Acceleration Mechanisms for Particle Injection and Power-law Formation during Transrelativistic Magnetic Reconnection}},
  journal       = {\apj},
  year          = 2020,
  month         = aug,
  volume        = {899},
  number        = {2},
  eid           = {151},
  pages         = {151},
  doi           = {10.3847/1538-4357/aba1e9},
  archiveprefix = {arXiv},
  eprint        = {2001.02732},
  primaryclass  = {astro-ph.HE},
  adsurl        = {https://ui.adsabs.harvard.edu/abs/2020ApJ...899..151K}
}

@article{2020ApJ...900..100R,
  author        = {{Ripperda}, Bart and {Bacchini}, Fabio and {Philippov}, Alexander A.},
  title         = {{Magnetic Reconnection and Hot Spot Formation in Black Hole Accretion Disks}},
  journal       = {\apj},
  year          = 2020,
  month         = sep,
  volume        = {900},
  number        = {2},
  eid           = {100},
  pages         = {100},
  doi           = {10.3847/1538-4357/ababab},
  archiveprefix = {arXiv},
  eprint        = {2003.04330},
  primaryclass  = {astro-ph.HE},
  adsurl        = {https://ui.adsabs.harvard.edu/abs/2020ApJ...900..100R}
}

@article{2021ApJ...922..261Z,
  author        = {{Zhang}, Hao and {Sironi}, Lorenzo and {Giannios}, Dimitrios},
  title         = {{Fast Particle Acceleration in Three-dimensional Relativistic Reconnection}},
  journal       = {\apj},
  year          = 2021,
  month         = dec,
  volume        = {922},
  number        = {2},
  eid           = {261},
  pages         = {261},
  doi           = {10.3847/1538-4357/ac2e08},
  archiveprefix = {arXiv},
  eprint        = {2105.00009},
  primaryclass  = {astro-ph.HE},
  adsurl        = {https://ui.adsabs.harvard.edu/abs/2021ApJ...922..261Z}
}

@article{2021MNRAS.506.1128S,
  author        = {{Sun}, Xiaochen and {Bai}, Xue-Ning},
  title         = {{Particle diffusion and acceleration in magnetorotational instability turbulence}},
  journal       = {\mnras},
  year          = 2021,
  month         = sep,
  volume        = {506},
  number        = {1},
  pages         = {1128-1147},
  doi           = {10.1093/mnras/stab1643},
  archiveprefix = {arXiv},
  eprint        = {2106.03098},
  primaryclass  = {astro-ph.HE},
  adsurl        = {https://ui.adsabs.harvard.edu/abs/2021MNRAS.506.1128S}
}

@article{2022ApJ...938...86B,
  author        = {{Bacchini}, Fabio and {Arzamasskiy}, Lev and {Zhdankin}, Vladimir and {Werner}, Gregory R. and {Begelman}, Mitchell C. and {Uzdensky}, Dmitri A.},
  title         = {{Fully Kinetic Shearing-box Simulations of Magnetorotational Turbulence in 2D and 3D. I. Pair Plasmas}},
  journal       = {\apj},
  year          = 2022,
  month         = oct,
  volume        = {938},
  number        = {1},
  eid           = {86},
  pages         = {86},
  doi           = {10.3847/1538-4357/ac8a94},
  archiveprefix = {arXiv},
  eprint        = {2206.07061},
  primaryclass  = {astro-ph.HE},
  adsurl        = {https://ui.adsabs.harvard.edu/abs/2022ApJ...938...86B}
}

@article{2022JPlPh..88a9014U,
  author        = {{Uzdensky}, Dmitri A.},
  title         = {{Relativistic non-thermal particle acceleration in two-dimensional collisionless magnetic reconnection}},
  journal       = {Journal of Plasma Physics},
  year          = 2022,
  month         = feb,
  volume        = {88},
  number        = {1},
  eid           = {905880114},
  pages         = {905880114},
  doi           = {10.1017/S0022377822000046},
  archiveprefix = {arXiv},
  eprint        = {2007.09533},
  primaryclass  = {astro-ph.HE},
  adsurl        = {https://ui.adsabs.harvard.edu/abs/2022JPlPh..88a9014U}
}

@article{2022MNRAS.514.1940D,
  author        = {{Das}, Santabrata and {Nandi}, Anuj and {Stalin}, C.~S. and {Rakshit}, Suvendu and {Dihingia}, Indu Kalpa and {Singh}, Swapnil and {Aktar}, Ramiz and {Mitra}, Samik},
  title         = {{On the origin of core radio emissions from black hole sources in the realm of relativistic shocked accretion flow}},
  journal       = {\mnras},
  year          = 2022,
  month         = aug,
  volume        = {514},
  number        = {2},
  pages         = {1940-1951},
  doi           = {10.1093/mnras/stac1398},
  archiveprefix = {arXiv},
  eprint        = {2205.07737},
  primaryclass  = {astro-ph.HE},
  adsurl        = {https://ui.adsabs.harvard.edu/abs/2022MNRAS.514.1940D}
}

@article{2022Sci...378..538I,
  author        = {{IceCube Collaboration} and {Abbasi}, R. and {Ackermann}, M. and {Adams}, J. and {Aguilar}, J.~A. and {Ahlers}, M. and {Ahrens}, M. and {Alameddine}, J.~M. and {Alispach}, C. and {Alves}, Jr., A.~A. and {Amin}, N.~M. and {Andeen}, K. and {Anderson}, T. and {Anton}, G. and {Arg{\"u}elles}, C. and {Ashida}, Y. and {Axani}, S. and {Bai}, X. and {Balagopal}, A.~V. and {Barbano}, V.~A. and {Barwick}, S.~W. and {Bastian}, B. and {Basu}, V. and {Baur}, S. and {Bay}, R. and {Beatty}, J.~J. and {Becker}, K. -H. and {Becker Tjus}, J. and {Bellenghi}, C. and {Benzvi}, S. and {Berley}, D. and {Bernardini}, E. and {Besson}, D.~Z. and {Binder}, G. and {Bindig}, D. and {Blaufuss}, E. and {Blot}, S. and {Boddenberg}, M. and {Bontempo}, F. and {Borowka}, J. and {B{\"o}ser}, S. and {Botner}, O. and {B{\"o}ttcher}, J. and {Bourbeau}, E. and {Bradascio}, F. and {Braun}, J. and {Brinson}, B. and {Bron}, S. and {Brostean-Kaiser}, J. and {Browne}, S. and {Burgman}, A. and {Burley}, R.~T. and {Busse}, R.~S. and {Campana}, M.~A. and {Carnie-Bronca}, E.~G. and {Chen}, C. and {Chen}, Z. and {Chirkin}, D. and {Choi}, K. and {Clark}, B.~A. and {Clark}, K. and {Classen}, L. and {Coleman}, A. and {Collin}, G.~H. and {Conrad}, J.~M. and {Coppin}, P. and {Correa}, P. and {Cowen}, D.~F. and {Cross}, R. and {Dappen}, C. and {Dave}, P. and {de Clercq}, C. and {Delaunay}, J.~J. and {Delgado L{\'o}pez}, D. and {Dembinski}, H. and {Deoskar}, K. and {Desai}, A. and {Desiati}, P. and {de Vries}, K.~D. and {de Wasseige}, G. and {de With}, M. and {Deyoung}, T. and {Diaz}, A. and {D{\'\i}az-V{\'e}lez}, J.~C. and {Dittmer}, M. and {Dujmovic}, H. and {Dunkman}, M. and {Duvernois}, M.~A. and {Dvorak}, E. and {Ehrhardt}, T. and {Eller}, P. and {Engel}, R. and {Erpenbeck}, H. and {Evans}, J. and {Evenson}, P.~A. and {Fan}, K.~L. and {Fazely}, A.~R. and {Fedynitch}, A. and {Feigl}, N. and {Fiedlschuster}, S. and {Fienberg}, A.~T. and {Filimonov}, K. and {Finley}, C. and {Fischer}, L. and {Fox}, D. and {Franckowiak}, A. and {Friedman}, E. and {Fritz}, A. and {F{\"u}rst}, P. and {Gaisser}, T.~K. and {Gallagher}, J. and {Ganster}, E. and {Garcia}, A. and {Garrappa}, S. and {Gerhardt}, L. and {Ghadimi}, A. and {Glaser}, C. and {Glauch}, T. and {Gl{\"u}senkamp}, T. and {Goldschmidt}, A. and {Gonzalez}, J.~G. and {Goswami}, S. and {Grant}, D. and {Gr{\'e}goire}, T. and {Griswold}, S. and {G{\"u}nther}, C. and {Gutjahr}, P. and {Haack}, C. and {Hallgren}, A. and {Halliday}, R. and {Halve}, L. and {Halzen}, F. and {Hanson}, M. Ha Minh K. and {Hardin}, J. and {Harnisch}, A.~A. and {Haungs}, A. and {Hebecker}, D. and {Helbing}, K. and {Henningsen}, F. and {Hettinger}, E.~C. and {Hickford}, S. and {Hignight}, J. and {Hill}, C. and {Hill}, G.~C. and {Hoffman}, K.~D. and {Hoffmann}, R. and {Hokanson-Fasig}, B. and {Hoshina}, K. and {Huang}, F. and {Huber}, M. and {Huber}, T. and {Hultqvist}, K. and {H{\"u}nnefeld}, M. and {Hussain}, R. and {Hymon}, K. and {in}, S. and {Iovine}, N. and {Ishihara}, A. and {Jansson}, M. and {Japaridze}, G.~S. and {Jeong}, M. and {Jin}, M. and {Jones}, B.~J.~P. and {Kang}, D. and {Kang}, W. and {Kang}, X. and {Kappes}, A. and {Kappesser}, D. and {Kardum}, L. and {Karg}, T. and {Karl}, M. and {Karle}, A. and {Katz}, U. and {Kauer}, M. and {Kellermann}, M. and {Kelley}, J.~L. and {Kheirandish}, A. and {Kin}, K. and {Kintscher}, T. and {Kiryluk}, J. and {Klein}, S.~R. and {Koirala}, R. and {Kolanoski}, H. and {Kontrimas}, T. and {K{\"o}pke}, L. and {Kopper}, C. and {Kopper}, S. and {Koskinen}, D.~J. and {Koundal}, P. and {Kovacevich}, M. and {Kowalski}, M. and {Kozynets}, T. and {Kun}, E. and {Kurahashi}, N. and {Lad}, N. and {Lagunas Gualda}, C. and {Lanfranchi}, J.~L. and {Larson}, M.~J. and {Lauber}, F. and {Lazar}, J.~P.},
  title         = {{Evidence for neutrino emission from the nearby active galaxy NGC 1068}},
  journal       = {Science},
  year          = 2022,
  month         = nov,
  volume        = {378},
  number        = {6619},
  pages         = {538-543},
  doi           = {10.1126/science.abg3395},
  archiveprefix = {arXiv},
  eprint        = {2211.09972},
  primaryclass  = {astro-ph.HE},
  adsurl        = {https://ui.adsabs.harvard.edu/abs/2022Sci...378..538I}
}

@article{2023ApJ...956L..36Z,
  author        = {{Zhang}, Hao and {Sironi}, Lorenzo and {Giannios}, Dimitrios and {Petropoulou}, Maria},
  title         = {{The Origin of Power-law Spectra in Relativistic Magnetic Reconnection}},
  journal       = {\apjl},
  year          = 2023,
  month         = oct,
  volume        = {956},
  number        = {2},
  eid           = {L36},
  pages         = {L36},
  doi           = {10.3847/2041-8213/acfe7c},
  archiveprefix = {arXiv},
  eprint        = {2302.12269},
  primaryclass  = {astro-ph.HE},
  adsurl        = {https://ui.adsabs.harvard.edu/abs/2023ApJ...956L..36Z}
}

@article{2023PhRvR...5d3023Z,
  author        = {{Zhdankin}, Vladimir and {Ripperda}, Bart and {Philippov}, Alexander A.},
  title         = {{Particle acceleration by magnetic Rayleigh-Taylor instability: Mechanism for flares in black hole accretion flows}},
  journal       = {Physical Review Research},
  year          = 2023,
  month         = oct,
  volume        = {5},
  number        = {4},
  eid           = {043023},
  pages         = {043023},
  doi           = {10.1103/PhysRevResearch.5.043023},
  archiveprefix = {arXiv},
  eprint        = {2302.05276},
  primaryclass  = {astro-ph.HE},
  adsurl        = {https://ui.adsabs.harvard.edu/abs/2023PhRvR...5d3023Z}
}

@article{2024ApJ...964L..26E,
  author   = {{Event Horizon Telescope Collaboration} and {Akiyama}, Kazunori and {Alberdi}, Antxon and {Alef}, Walter and {Algaba}, Juan Carlos and {Anantua}, Richard and {Asada}, Keiichi and {Azulay}, Rebecca and {Bach}, Uwe and {Baczko}, Anne-Kathrin and {Ball}, David and {Balokovi{\'c}}, Mislav and {Bandyopadhyay}, Bidisha and {Barrett}, John and {Baub{\"o}ck}, Michi and {Benson}, Bradford A. and {Bintley}, Dan and {Blackburn}, Lindy and {Blundell}, Raymond and {Bouman}, Katherine L. and {Bower}, Geoffrey C. and {Boyce}, Hope and {Bremer}, Michael and {Brinkerink}, Christiaan D. and {Brissenden}, Roger and {Britzen}, Silke and {Broderick}, Avery E. and {Broguiere}, Dominique and {Bronzwaer}, Thomas and {Bustamante}, Sandra and {Byun}, Do-Young and {Carlstrom}, John E. and {Ceccobello}, Chiara and {Chael}, Andrew and {Chan}, Chi-kwan and {Chang}, Dominic O. and {Chatterjee}, Koushik and {Chatterjee}, Shami and {Chen}, Ming-Tang and {Chen}, Yongjun and {Cheng}, Xiaopeng and {Cho}, Ilje and {Christian}, Pierre and {Conroy}, Nicholas S. and {Conway}, John E. and {Cordes}, James M. and {Crawford}, Thomas M. and {Crew}, Geoffrey B. and {Cruz-Osorio}, Alejandro and {Cui}, Yuzhu and {Dahale}, Rohan and {Davelaar}, Jordy and {De Laurentis}, Mariafelicia and {Deane}, Roger and {Dempsey}, Jessica and {Desvignes}, Gregory and {Dexter}, Jason and {Dhruv}, Vedant and {Dihingia}, Indu K. and {Doeleman}, Sheperd S. and {Dougall}, Sean and {Dzib}, Sergio A. and {Eatough}, Ralph P. and {Emami}, Razieh and {Falcke}, Heino and {Farah}, Joseph and {Fish}, Vincent L. and {Fomalont}, Edward and {Ford}, H. Alyson and {Foschi}, Marianna and {Fraga-Encinas}, Raquel and {Freeman}, William T. and {Friberg}, Per and {Fromm}, Christian M. and {Fuentes}, Antonio and {Galison}, Peter and {Gammie}, Charles F. and {Garc{\'\i}a}, Roberto and {Gentaz}, Olivier and {Georgiev}, Boris and {Goddi}, Ciriaco and {Gold}, Roman and {G{\'o}mez-Ruiz}, Arturo I. and {G{\'o}mez}, Jos{\'e} L. and {Gu}, Minfeng and {Gurwell}, Mark and {Hada}, Kazuhiro and {Haggard}, Daryl and {Haworth}, Kari and {Hecht}, Michael H. and {Hesper}, Ronald and {Heumann}, Dirk and {Ho}, Luis C. and {Ho}, Paul and {Honma}, Mareki and {Huang}, Chih-Wei L. and {Huang}, Lei and {Hughes}, David H. and {Ikeda}, Shiro and {Impellizzeri}, C.~M. Violette and {Inoue}, Makoto and {Issaoun}, Sara and {James}, David J. and {Jannuzi}, Buell T. and {Janssen}, Michael and {Jeter}, Britton and {Jiang}, Wu and {Jim{\'e}nez-Rosales}, Alejandra and {Johnson}, Michael D. and {Jorstad}, Svetlana and {Joshi}, Abhishek V. and {Jung}, Taehyun and {Karami}, Mansour and {Karuppusamy}, Ramesh and {Kawashima}, Tomohisa and {Keating}, Garrett K. and {Kettenis}, Mark and {Kim}, Dong-Jin and {Kim}, Jae-Young and {Kim}, Jongsoo and {Kim}, Junhan and {Kino}, Motoki and {Koay}, Jun Yi and {Kocherlakota}, Prashant and {Kofuji}, Yutaro and {Koch}, Patrick M. and {Koyama}, Shoko and {Kramer}, Carsten and {Kramer}, Joana A. and {Kramer}, Michael and {Krichbaum}, Thomas P. and {Kuo}, Cheng-Yu and {La Bella}, Noemi and {Lauer}, Tod R. and {Lee}, Daeyoung and {Lee}, Sang-Sung and {Leung}, Po Kin and {Levis}, Aviad and {Li}, Zhiyuan and {Lico}, Rocco and {Lindahl}, Greg and {Lindqvist}, Michael and {Lisakov}, Mikhail and {Liu}, Jun and {Liu}, Kuo and {Liuzzo}, Elisabetta and {Lo}, Wen-Ping and {Lobanov}, Andrei P. and {Loinard}, Laurent and {Lonsdale}, Colin J. and {Lowitz}, Amy E. and {Lu}, Ru-Sen and {MacDonald}, Nicholas R. and {Mao}, Jirong and {Marchili}, Nicola and {Markoff}, Sera and {Marrone}, Daniel P. and {Marscher}, Alan P. and {Mart{\'\i}-Vidal}, Iv{\'a}n and {Matsushita}, Satoki and {Matthews}, Lynn D. and {Medeiros}, Lia and {Menten}, Karl M. and {Michalik}, Daniel and {Mizuno}, Izumi and {Mizuno}, Yosuke and {Moran}, James M. and {Moriyama}, Kotaro and {Moscibrodzka}, Monika and {Mulaudzi}, Wanga and {M{\"u}ller}, Cornelia and {M{\"u}ller}, Hendrik and {Mus}, Alejandro and {Musoke}, Gibwa and {Myserlis}, Ioannis and {Nadolski}, Andrew and {Nagai}, Hiroshi and {Nagar}, Neil M. and {Nakamura}, Masanori and {Narayanan}, Gopal and {Natarajan}, Iniyan and {Nathanail}, Antonios and {Fuentes}, Santiago Navarro and {Neilsen}, Joey and {Neri}, Roberto and {Ni}, Chunchong and {Noutsos}, Aristeidis and {Nowak}, Michael A. and {Oh}, Junghwan and {Okino}, Hiroki and {Olivares}, H{\'e}ctor and {Ortiz-Le{\'o}n}, Gisela N. and {Oyama}, Tomoaki and {{\"O}zel}, Feryal and {Palumbo}, Daniel C.~M. and {Paraschos}, Georgios Filippos and {Park}, Jongho and {Parsons}, Harriet and {Patel}, Nimesh and {Pen}, Ue-Li},
  title    = {{First Sagittarius A* Event Horizon Telescope Results. VIII. Physical Interpretation of the Polarized Ring}},
  journal  = {\apjl},
  year     = 2024,
  month    = apr,
  volume   = {964},
  number   = {2},
  eid      = {L26},
  pages    = {L26},
  doi      = {10.3847/2041-8213/ad2df1},
  adsurl   = {https://ui.adsabs.harvard.edu/abs/2024ApJ...964L..26E}
}

@article{2024MNRAS.530.1866S,
  author        = {{Sandoval}, Astor and {Riquelme}, Mario and {Spitkovsky}, Anatoly and {Bacchini}, Fabio},
  title         = {{Particle-in-cell simulations of the magnetorotational instability in stratified shearing boxes}},
  journal       = {\mnras},
  year          = 2024,
  month         = may,
  volume        = {530},
  number        = {2},
  pages         = {1866-1884},
  doi           = {10.1093/mnras/stae959},
  archiveprefix = {arXiv},
  eprint        = {2308.12348},
  primaryclass  = {astro-ph.HE},
  adsurl        = {https://ui.adsabs.harvard.edu/abs/2024MNRAS.530.1866S}
}

@article{2024PhRvL.133d5202B,
  author        = {{Bacchini}, Fabio and {Zhdankin}, Vladimir and {Gorbunov}, Evgeny A. and {Werner}, Gregory R. and {Arzamasskiy}, Lev and {Begelman}, Mitchell C. and {Uzdensky}, Dmitri A.},
  title         = {{Collisionless Magnetorotational Turbulence in Pair Plasmas: Steady-State Dynamics, Particle Acceleration, and Radiative Cooling}},
  journal       = {\prl},
  year          = 2024,
  month         = jul,
  volume        = {133},
  number        = {4},
  eid           = {045202},
  pages         = {045202},
  doi           = {10.1103/PhysRevLett.133.045202},
  archiveprefix = {arXiv},
  eprint        = {2401.01399},
  primaryclass  = {astro-ph.HE},
  adsurl        = {https://ui.adsabs.harvard.edu/abs/2024PhRvL.133d5202B}
}

@article{2025ApJ...980..255Y,
  author        = {{Yang}, Qi-Rui and {Liu}, Ruo-Yu and {Wang}, Xiang-Yu},
  title         = {{Could the Neutrino Emission of TXS 0506+056 Come from the Accretion Flow of the Supermassive Black Hole?}},
  journal       = {\apj},
  year          = 2025,
  month         = feb,
  volume        = {980},
  number        = {2},
  eid           = {255},
  pages         = {255},
  doi           = {10.3847/1538-4357/adaea4},
  archiveprefix = {arXiv},
  eprint        = {2411.17632},
  primaryclass  = {astro-ph.HE},
  adsurl        = {https://ui.adsabs.harvard.edu/abs/2025ApJ...980..255Y}
}

@article{2025ApJ...982L..28G,
  author        = {{Gorbunov}, Evgeny A. and {Bacchini}, Fabio and {Zhdankin}, Vladimir and {Werner}, Gregory R. and {Begelman}, Mitchell C. and {Uzdensky}, Dmitri A.},
  title         = {{First-principles Measurement of Ion and Electron Energization in Collisionless Accretion Flows}},
  journal       = {\apjl},
  year          = 2025,
  month         = mar,
  volume        = {982},
  number        = {1},
  eid           = {L28},
  pages         = {L28},
  doi           = {10.3847/2041-8213/adbca4},
  archiveprefix = {arXiv},
  eprint        = {2410.02872},
  primaryclass  = {astro-ph.HE},
  adsurl        = {https://ui.adsabs.harvard.edu/abs/2025ApJ...982L..28G}
}

@article{buneman1993computer,
  title   = {Computer space plasma physics},
  author  = {Buneman, O},
  journal = {Terra Scientific, Tokyo},
  volume  = {67},
  year    = {1993}
}
\bibliographystyle{aasjournalv7.1}



\end{document}